   \documentclass[reprint,aps]{revtex4-1}
                   \usepackage{dcolumn}
\usepackage{amsmath}
\usepackage{graphicx}
\usepackage{bm}
\usepackage{color}
\usepackage{comment}

\newcommand{\GeV}{\makebox{ GeV}}
\newcommand{\beq}{\begin{equation}}
\newcommand{\enq}{\end{equation}}
\newcommand{\beqa}{\begin{eqnarray}}
\newcommand{\beqast}{\begin{eqnarray*}}
\newcommand{\enqa}{\end{eqnarray}}
\newcommand{\enqast}{\end{eqnarray*}}

\def\GeV{\nobreak\,\mbox{GeV}}

\providecommand{\U}[1]{\protect\rule{.1in}{.1in}}

\begin{document}

\title{   Elastic amplitudes studied with the LHC  measurements  at 7 and 8 TeV }

\author{A. K. Kohara $^{\rm a,~c~ \ast}$     
\thanks{Corresponding author. Email: kendi@if.ufrj.br\vspace{6pt}} }
\author{E. Ferreira $^{\rm a  } $ }
\author{T. Kodama $^{\rm a,~b}$   }
\author{M. Rangel $^{\rm a}$     }

\begin{abstract}
Recent  measurements of the differential cross sections in the forward region of
 pp elastic scattering at 7 and  8 TeV   show precise form of the $t$ dependence. 
  We propose a detailed analysis of these measurements including the structures 
of the real and imaginary parts of the scattering amplitude.  A good description 
is achieved, confirming in all experiments the existence of a zero in the real 
part  in the forward region close to the origin, in agreement with the prediction 
of a theorem by A. Martin, with important role in the observed form  of $d\sigma/dt$. 
 Universal value for the position of this zero and regularity in other features 
of the amplitudes are found, leading to quantitative  predictions for the forward 
elastic scattering at 13 TeV.
\end{abstract}

\affiliation{ $ ^{\rm a}$ {\em Instituto de F\'{\i}sica, Universidade Federal do Rio de Janeiro \\
C.P. 68528, Rio de Janeiro 21945-970, RJ, Brazil }    }
\affiliation{ $ ^{\rm b}$ {\em Instituto de F\'{\i}sica, Universidade Federal Fluminense \\
C.P. , Niter\'oi  24210-346, RJ, Brazil }    }
\affiliation{ $ ^{\rm c}$ {\em Centro Brasileiro de Pesquisas F\'isicas - CBPF \\
C.P. , Rio de Janeiro , RJ, Brazil }    }

\maketitle

\section {LHC experiments in elastic ${\rm pp}$  scattering}

With an enormous  gap in the center-of-mass energy  with respect to previous  
data in pp and $\rm p \bar p$  
 scattering, the Totem and Atlas experimental groups  at  LHC  have recently  measured 
$d\sigma/dt$   in forward $t$ ranges at $\sqrt{s}$ = 7 and 8 TeV \cite{T7,A7,T8,A8}. 
These measurements offer a unique opportunity to investigate 
the behaviour of p-p collisions at the highest energies reached in 
laboratory. A detailed and precise analysis of these data can establish 
a precious milestone for the understanding of the high energy behaviour of  p-p interactions.
The datasets and  their $t$ ranges   are listed   in Table \ref{datasets}, 
where we use obvious    notation T7, T8, A7, A8 to specify  Totem (T) and Atlas (A) 
Collaborations and center-of-mass energies 7 and 8 TeV. 
\begin{table*}[t]
\begin{center}
 \vspace{0.5cm}
\begin{tabular}{ c c c c c c c c c}
\hline
\hline
      &             &               &   &  &  &   &         \\
\hline 
  $\sqrt{s}$  & dataset  & $\Delta {|t|}$  range &  N         & Ref. &  $\sigma$ & $B_I$& $ \rho$        \\
     (GeV)    &               & $(\GeV^{2})$          &  points    &      &   (mb)    & ( GeV$^{-2})$ &     \\
\hline 
  7           &  Totem T7       &  0.005149-0.3709      & 87         & 1    & 98.6$\pm$2.2      &19.9$\pm$0.3     & 0.14 (fix)$^{\rm a}$   & \\
\hline
   7          &  Atlas A7       &  0.0062-0.3636        & 40        & 2     & $ 95.35\pm  0.38 $  & $19.73\pm 0.14 $ & 0.14 (fix) $^{\rm b}$    &    \\
 \hline
   8           &  Totem T8       & 0.000741-0.19478      & 60       & 3     &  $103.0 \pm 2.3   $  &  19.56 $\pm$ 0.13     & (0.12 $\pm$ 0.03) $^{\rm c}$   &   \\
\hline
   8          &  Atlas A8       &   0.0105-0.3635       & 39        & 4     & $96.07\pm 0.18 $    &  $19.74\pm 0.05$     &  0.1362  (fix)$^{\rm d}$ &    \\
 \hline
 \end{tabular}
 \caption{ Measurements at $\sqrt{s}=$7 and 8 TeV from Totem and Atlas Collaborations at 
LHC \cite{T7,A7,T8,A8}. The $\rho$ values for $^{\rm a}$ $^{\rm b}$  are taken from COMPETE Collaboration \cite{COMPETE} . For $^{\rm c}$ the $\rho$ value was obtained by the authors with 
a forward SET-I and kept fixed in a complete  SET-II as explained in the sub-section with analysis of T8.  For $^{\rm d}$ the $\rho$ value  is taken from \cite{PDG}.}  
\label{datasets}
\end{center}
\end{table*}

In order to  build a bridge towards theoretical models aiming at the understanding 
of the dynamics, it is important that   the analysis of these LHC data be made 
with identification  of the structures of the individual parts of the complex scattering  amplitude. 
The disentanglement of the two terms in the observed  modulus $d\sigma/dt$  is the 
crucial task. At each energy, parameterizations must search to exhibit clearly  the 
properties of magnitudes, signs, slopes  and zeros of the real and imaginary parts.
External support, as dispersion relations and connections with analyses at other 
energies, give  important   clues.  
  The intervention of the electromagnetic interactions must be treated coherently 
with a proposed  analytical form for the nuclear part, and account must be taken of 
phase of the Coulomb-Nuclear Interference (CNI), which is calculated in Appendix A.

 In the present work we   perform a detailed 
examination of the data trying to satisfy these requirements.
 Each part of the amplitude is written with  an exponential factor with a slope, 
multiplying a linear   term in $t$, thus with three parameters.  
These analytical forms are sufficient to describe  the properties of the 
nuclear parts.    The six parameters for each dataset  
are studied  using fits to data with 
appropriate statistical control. Correlations are studied, and  resulting values 
are proposed for each dataset.  
Good description of the measurements is obtained, with details in the shape 
of the forward diffractive peak,  exhibiting the zero of the real part 
predicted in the theorem by  A. Martin  \cite{Martin}, and with observation 
in the forward range of the ingredients that construct the imaginary zero 
responsible for the dip in $d\sigma/dt$ observed when data exist at higher $|t|$. 
 
According to  analyses and models of pp and ${\rm p \bar p} $ scattering at high energies 
   \cite{LHC7TeV,LHC8TeV,Models}, including full-$t$ ranges, 
the imaginary part has a zero near the marked dip observed in $d\sigma/dt$ at about 0.4-0.6 $\GeV^2$, 
while the real part  starts positive at $t=0$, has a zero  for small $|t|$ (Martin's  theorem), and a 
second zero after  the dip. The descriptions of the models  differ for higher $|t|$ about the position 
of this second real zero and about the  existence  of further imaginary or real zeros.    
Models describing large $|t|$ ranges   have  different motivations and dynamical structures, 
and may be analytically very sophisticated, trying to represent the observed 
shapes of dip, bump, tail, in the angular  dependence. 
 However in the forward range the analytical structure required to describe 
data may be very simple. In the present work we show that real and imaginary parts including exponential slopes and linear factors,
combined with the    electromagnetic interference,  contain 
the essential ingredients for a  precise representation. The second real zero occurs outside the studied range and does not influence the analysis. 
Particularly due to the small magnitude of the $\rho$ parameter, 
 the disentanglement  of the two parts is   not trivial, requiring careful analysis, 
 and  still leaving room for some subjective but physically reasonable choice.
Once the amplitudes are identified, the results  provide a necessary connection 
 between data and 
theoretical models of microscopic nature for the strong interaction dynamics. 

The deviation from the pure exponential form in 
  $d\sigma/dt$ is obvious beforehand, 
since $d\sigma/dt$   is a sum of two independent squares. More clarity 
can be obtained in the analysis  with  the identification of the 
two parts of the complex amplitude and their control by fundamental 
constraints (dispersion relations, Martin's theorem for the real part, zero 
of the imaginary part anticipating the dip).

Of the four datasets, T8 is the only one reaching very small $|t|$, allowing the 
more complete investigation of some details, such as the influence of the Coulomb phase.  
However a comparative analysis of the four cases is extremely important, since   
 observing coherence in some characteristics  we may believe that 
there is reliability in the descriptions. The energy dependence of pp elastic scattering
is very smooth, and features of 7 and 8 TeV data  must support each other
in a unified treatment.

 A theorem  by Andr\'e  Martin proves that the real part has a zero close 
to the origin \cite{Martin}. The abstract of the paper says, literally  
\bigskip  

{  \it  We show that if for fixed  negative (physical) square of the momentum 
transfer  $t$, the   differential cross section $d\sigma/dt$  tends to zero and 
if the total cross section  tends to infinity, when the energy goes to infinity, 
the real part of the even signature amplitude 
cannot have a constant sign near $t=0$. @ 1997 Elsevier Science B. V. }

\bigskip

Thus the real part in pp and p\=p scattering  has a zero close to the origin, with location
approaching  $t=0$ as the energy increases. This constraint has been confirmed  in 
previous analyses at LHC and lower energies \cite{LHC7TeV,LHC8TeV}, with the conclusion that 
the position of the  first zero of the real part 
behaves like $ |t_R| ~= A+ 1/(c_0 +c_1 ~\log{s}+ c_2 ~\log^2{s}). $   
 
Although the analytical properties of the  amplitude 
are defined in   t-space, the insights for the construction of  theoretical models 
are   natural in the geometrical space, where the physical intuition to 
build amplitudes may be  represented. The Fourier transformed space is 
appropriate to study asymptotic  properties of the cross sections such as   
questioning whether the proton behaves as a black or a gray disk. The profile 
functions are also  convenient tools to study the unitarity constraint. 
Of course we recognize that the amplitudes written for the short forward 
$t$-range cannot lead to a sufficient understanding of $b$-space properties. 
 Even thus, we believe that the relationship is important, and the Fourier 
transformation of our amplitudes to $b$-space is analytically performed and 
its properties discussed  in Appendix B.

  This paper is organized as follows. Section II describes the formalism of the 
proposed model; Section III presents the results of the model fits to the four 
LHC measurements; Section IV summarizes the numerical analysis ; Section V 
concludes the paper.

\section{Amplitudes and Observables in Forward Scattering  \label{formalism}}

 In the analysis limited to the forward ranges shown in 
Table \ref{datasets}, 
 the  expectations are satisfied 
writing the differential cross section in the form 
\begin{eqnarray}   
&&\frac{d\sigma}{dt}=\pi\left(\hbar c\right)^{2}  \times           \nonumber \\
&&  \Big\{\Big[\frac{\sigma(\rho-\mu_R t) }{4\pi\left(\hbar c\right)^{2}}~{{e}^{B_{R}t/2}
+F^{C}(t)\cos{(\alpha\phi)}\Big]^{2}}\nonumber\\
&& +\Big[\frac{\sigma (1-\mu_I t)}{4\pi\left(  \hbar c\right)  ^{2}}~{{e}^{B_{I}t/2}%
+F^{C}(t)\sin{(\alpha\phi)}\Big]^{2}\Big\}~,}   
  \label{diffcross_eq}
 \end{eqnarray}
where $t\equiv-|t|$,  $\alpha$ is the fine structure constant and 
$(\hbar c)^2=0.3894 $ mb $\GeV^2$. 
   This expression is applied for 
pp an p\=p, and  the parameters are specific for each case.
$F^C(t)$ and $\alpha\phi(t)$ represent the form factor and phase of the Coulomb interaction. 
The phase $\alpha\phi$ has opposite signs for pp and p\=p sccattering.

  The real electromagnetic amplitude is  given 
in terms of the proton form factor
\begin{equation}
F^{C}(t)~=(-/+)~\frac{2\alpha }{|t|}~F_{\mathrm{proton}}^{2}(t)~,
\label{coulomb}
\end{equation}%
for the pp$/$p$\mathrm{{\bar{p}}}$ collisions. The proton form factor is
taken as%
\begin{equation}
F_{\mathrm{proton}}(t)=[\Lambda^2/(\Lambda^2+|t|)]^{2}~,  \label{ff_proton}
\end{equation}%
where $\Lambda^2=0.71\GeV^{2}$. 
The phase of the Coulomb-Nuclear interference is discussed in  Appendix A.

We have thus assumed the  imaginary amplitude with 
  an exponential that accounts for the forward diffractive peak and a linear factor 
that accounts for the zero that occurs near the dip in $d\sigma/dt$, writing  
the simple  form  
   \begin{equation}
T_I^N (t)= \frac{1}{4 \sqrt{\pi}\left(\hbar c\right)^2 } ~ \sigma ( 1 -\mu_I t )~ e^{B_I t/2} ~ 
\label{imag_TI}
\end{equation}
and 
 \begin{equation}
T_I(t)=T_I^N (t)+\sqrt{\pi} F^C(t) \sin(\alpha \phi) ~. 
\label{imag_TI_C}
\end{equation}
 The influence of the parameter   $\mu_I$ 
depends on the range of the data analyzed.
To include the   influence of the first real  zero, we write 
\begin{equation}
    T_R^N(t)= \frac{1}{4 \sqrt{\pi} \left(\hbar c\right)^2} ~ \sigma (\rho -\mu_R t) ~ e^{B_R t/2}  
\label{real_TR}
\end{equation}
and 
\begin{equation}
 T_R(t)=T_R^N(t)+\sqrt{\pi} F^C(t) \cos(\alpha \phi) ~. 
\label{real_TR_C}
\end{equation}
In order to check the influence of 
a second zero in the real part, we could add in the amplitude a term $\xi_R~ t^2$, 
but actually it has no  effect in the present analysis.  

The  normalization  is defined by
\begin{equation}
  \sigma(s)= 4\sqrt{\pi}(\hbar  c)^2 ~T_I^N(s,t=0)  
\end{equation} 
and for the pure nuclear interaction
 \begin{equation}
  \frac{d\sigma}{dt}= (\hbar  c)^2 [(T_I^N)^2+(T_R^N)^2] ~.   
\end{equation}

At $t=0$, we have  the usual definition of the $\rho$ parameter 
\begin{equation}
\rho = \frac{T_R^N(0)}{T_I^N(0)} . 
\label{real_forw0}
\end{equation}
  
With positive  $\rho$ and negative $\mu_R$ (this is what we have in pp at high energies,
as our analysis shows), 
there is a zero in the real 
amplitude, namely Martin's zero, located at 
\begin{equation}
 t_R = \frac{\rho}{\mu_R}~. 
\label{tR_position}
\end{equation}
 The position of this zero and the magnitudes of the real and imaginary amplitudes in its neighborhood  
are responsible for details in the deviation of the differential cross section from a pure 
exponential behaviour. 

The derivatives of the nuclear amplitudes at $t=0$ are 
\begin{equation} 
 \frac{d}{dt}\log T_I^N (t)\bigg|_0= \frac{1}{2}[B_I-2 \mu_I]= \frac{1}{2} B_I^{eff}
\label{BI_eff}
 \end{equation} 
and 
\begin{equation} 
\frac{d}{dt} \log T_R^N (t)\bigg|_0= \frac{1}{2}[B_R-2 \frac{\mu_R}{\rho}]= \frac{1}{2} B_R^{eff} ~. 
\label{BR_eff}
 \end{equation} 
The average slope measured directly in $d\sigma/dt$  is the quantity 
\begin{equation}
  B= \frac{2}{(d\sigma/dt)\big|_0} [\frac{d}{dt}(d\sigma/dt)]\bigg|_0=  
  \frac{1}{1+\rho^2} \big [B_I^{eff}+\rho^2 B_R^{eff}\big]  ~. 
\label{slope_dsigmadt}
\end{equation}

We remark that parameters are determined fitting data in limited $|t|$ ranges, at finite 
distance from the origin, so that the values obtained  depend on the 
analytical forms  (\ref{imag_TI},\ref{imag_TI_C},\ref{real_TR},\ref{real_TR_C})  
of the amplitudes. 
In particular, the slope parameter usually written in   form 
  $  d\sigma/dt= \sigma (\rho+i) \exp(Bt)$ 
does not agree with the expression for the differential cross section as
sum of  two independent squared  magnitudes, each with its own slope. The assumption 
that $B_R$ 
and $B_I$ are equal is not justified. The  average  quantity  $B$ alone gives  rough and 
unsatisfactory information.  
The importance of the different slopes in the analysis of pp elastic scattering has 
been investigated in the framework of the so called dispersion relations for slopes \cite{EF2007}. 
It is important to note that also   the Coulomb-Nuclear 
phase $\phi(t)$ depends essentially on the form of the  nuclear amplitudes \cite{KL}.  In 
the Appendix A, generalizing previous work \cite{WY,Cahn,LHC7TeV}, we derive the 
expression for the phase to be used with the  assumed amplitudes written above. 

Of course the six parameters are correlated, and in the present work we investigate the 
bounds of the correlations. We attempt  to identify  values 
of parameters that may be considered as common representatives for different  
measurements. We show  that the  differences between the two experimental collaborations 
may  be restricted  to  quantities characterizing normalization.  
 The question of normalization is essential, and our inputs are  the values
of $d\sigma/dt$ given in the experimental papers  \cite{T7,A7,T8,A8}. 

The extraction of forward parameters in pp scattering  has difficulties due 
to the small value of the $\rho$ parameter, and consequently  has suffered in many 
analyses  from neglect of the properties of the real part.   
In our view the values of $\sigma$, $\rho$, $B$ appearing in universal databases
\cite{PDG,COMPETE} 
as if they were direct experimental measurements should give room for 
critically controlled  phenomenological determinations. 
A proper consideration for the properties of the complex amplitude is necessary. 
We observe 
that the properties that $B_R \neq B_I$ and of the presence of zeros are common to 
several models  \cite{LHC7TeV,LHC8TeV,Models}. The determination 
of the amplitudes for all $|t|$ in several models is coherent. 

We observe that the polynomial factors written in the exponent in some 
parameterizations of data \cite{T8}
correspond to  the  linear and quadratic  factors mentioned 
above, if the assumption is made that they are much smaller than 1  and 
can be converted into exponentials. 
However  this substitution is not convenient, because it does not show explicitly the 
essential zeros, and it also gives unsatisfactory parameterization that cannot 
be extended even to nearby $|t|$ values.

 We thus have the framework necessary   for the analysis of the data,
    with clear identification of the role of the free  parameters.
  The    quantities to be determined for each dataset  are $\sigma$, $\rho$, 
 $B_I$, $B_R$, $\mu_I$, $\mu_R$. 

 \clearpage

\section{Data analysis}
 

\begin{table*}[t]
\begin{center}
 \vspace{0.5cm}
 \small
\begin{tabular}{c c  c c c c c c c c c c}
\hline 
\hline 
& N & $\phi$  &$\sigma$ & $\rho$        &$ B_{\rm I}  $      &$ B_{\rm R}  $      & $~\mu_{\rm R}  $     & $ \mu_{\rm I} $  & $t_R$  & $B_I^{\rm eff}$ &  $\chi^2/$ndf\\
  &   &      & (mb)                &  &$(\GeV^{-2})$ &$(\GeV^{-2})$ & $ (\GeV^{-2})$& $ (\GeV^{-2})$ & $(\GeV^{2})$ & $(\GeV^{-2})$ &         \\
\hline
\hline
      &     &     &   & &     &         &   &  &  &   &         \\
 
  \multicolumn{11}{c}{Condition I) - all six  parameters free  }         \\
\hline
T8 & 60 & ff    &102.70$\pm$0.07& 0.09$\pm$0.01& 19.96$\pm $0.05      &23.55$\pm$0.39      & -2.78$\pm$0.12     & -0.001$\pm$0.023 & -0.032$\pm$0.004& 19.96$\pm$0.07 & 58.53/54  \\
\hline
T8& 60 & 0   &102.60$\pm$0.09& 0.11$\pm$0.01& 19.77$\pm$0.05      &23.81$\pm$0.46      & -2.78$\pm$0.14     &-0.049$\pm$0.023 & -0.040$\pm$0.004& 19.87$\pm$0.07  &58.89/54  \\
 \hline
A8&39& ff    &97.02$\pm$0.52&0.11$\pm$0.08& 16.50$\pm$0.94      &22.76$\pm$1.22      & -3.37$\pm$0.24     & -1.65$\pm$0.41& -0.034$\pm$0.024& 19.80$\pm$1.25& 27.62/33  \\
 \hline
A8&39& 0   &96.91$\pm$0.60& 0.13$\pm$0.08& 16.42$\pm$0.94      &22.71$\pm$1.26      & -3.33$\pm$0.24     & -1.65$\pm$0.41 & -0.037$\pm$0.024& 19.72$\pm$1.24 &27.51/33  \\
\hline
T7& 87 & ff     &99.97$\pm$0.36 &0.11$\pm$0.06& 15.25$\pm$0.75      &23.64$\pm$0.61      & -4.53$\pm$0.37     & -2.45$\pm$0.20 & -0.024$\pm$0.013 &  20.15$\pm$0.85 & 93.48/81  \\
\hline
T7 & 87& 0   &99.82$\pm$0.43& 0.13$\pm$0.06& 15.10$\pm$0.72      &23.48$\pm$0.60      & -4.45$\pm$0.36     & -2.46$\pm$0.19& -0.029$\pm$0.014 & 20.02$\pm$0.81 & 93.52/81  \\
 \hline
A7&40&ff     &95.77$\pm$0.16&0.00$\pm$0.04 & 15.55$\pm$1.06      &21.74$\pm$0.49      & -4.03$\pm$0.65     & -2.36$\pm$0.58 & 0.000$\pm$0.010 & 20.27$\pm$1.57 & 20.93/34  \\
 \hline
A7 &40&0    &95.70$\pm$0.16&0.00$\pm$0.13& 15.58$\pm$1.18      &21.66$\pm$0.54      & -3.92$\pm$0.69     & -2.31$\pm$0.64 & 0.000$\pm$0.033 & 20.20$\pm$1.74&  20.58/34  \\
\hline
\hline
      &     &     &   & &    &          &   &  &  &   &         \\
 
   \multicolumn{11}{c}{Condition II) - $\rho = 0.14$ fixed  (suggested by dispersion relations) }         \\
      \hline
      \hline
T8& 60& ff     &102.60$\pm$0.16& 0.14 (fix)& 19.85$\pm$0.14      &24.77$\pm$0.57      & -3.07$\pm$0.26     & -0.003$\pm$0.274 &  -0.046$\pm$0.004 &19.86$\pm$0.56 & 66.35/55  \\
\hline
T8& 60& 0    &102.50$\pm$0.16& 0.14 (fix)& 19.77$\pm$0.57      &24.55$\pm$0.59      & -2.93$\pm$0.26     & -0.02$\pm$0.86 & -0.048$\pm$0.004& 19.81$\pm$1.81& 61.29/55  \\
 \hline
A8&39 & ff  &96.83$\pm$0.13&0.14 (fix) & 16.40$\pm$0.95      &22.83$\pm$1.40      & -3.33$\pm$0.22     & -1.63$\pm$0.43 & -0.042$\pm$0.003& 19.66$\pm$1.28 & 27.71/34  \\
 \hline
A8&39 & 0  &96.80$\pm$0.13&0.14 (fix)& 16.36$\pm$0.92      &22.74$\pm$1.35      & -3.31$\pm$0.22     & -1.64$\pm$0.42 &-0.042$\pm$0.003 & 19.64$\pm$1.25 & 27.54/34  \\
\hline
T7 & 87& ff    &99.83$\pm$0.20 &0.14 (fix) & 16.11$\pm$0.13      &25.00$\pm$1.77      & -4.21$\pm$0.33     & -1.97$\pm$0.10 & -0.033$\pm$0.003 &20.05$\pm$0.24 & 95.02/82  \\
\hline
T7& 87& 0     &99.75$\pm$0.20& 0.14 (fix) & 16.10$\pm$0.13      &25.00$\pm$1.88      & -4.13$\pm$0.33     & -1.95$\pm$0.09 & -0.034$\pm$0.003& 20.00$\pm$0.22 & 94.94/82  \\
\hline
A7 &40 &ff   &95.80$\pm$0.17&0.14 (fix) & 14.15$\pm$0.47      &21.45$\pm$0.34      & -4.54$\pm$0.39     & -2.67$\pm$0.24 & -0.031$\pm$0.003 &19.49$\pm$0.67 & 25.95/35  \\
 \hline
A7 &40&0   &95.73$\pm$0.17&0.14 (fix) & 14.15$\pm$0.49      &21.37$\pm$0.35      & -4.47$\pm$0.39     & -2.65$\pm$0.25 & -0.031$\pm$0.003& 19.45$\pm$0.70 & 25.11/35  \\
\hline
\hline
      &     &    & &   & &              &   &  &  &   &         \\
 
   \multicolumn{11}{c}{Condition III) - $\mu_I = - 2.16 \GeV^{-2}$ \cite{LHC8TeV} or $-2.14   \GeV^{-2} $ \cite{LHC7TeV}    }      \\
      \hline
      \hline
T8 & 60& ff    &102.60$\pm$0.14& 0.08$\pm$0.02& 15.50$\pm$0.11      &20.79$\pm$0.44      & -3.47$\pm$0.19     & -2.16 (fix) & -0.023$\pm$0.006& 19.82$\pm$0.11 & 58.98/55  \\
\hline
T8& 60& 0     &102.50$\pm$0.16& 0.11$\pm$0.02 & 15.36$\pm$0.11      &20.76$\pm$0.43      & -3.49$\pm$0.18     & -2.16 (fix) & -0.032$\pm$0.006 &19.68$\pm$0.11&  59.55/55  \\
 \hline
A8&39 & ff  &96.88$\pm$0.62 & 0.13$\pm$0.07  & 15.30$\pm$0.47      &21.68$\pm$0.36      & -3.69$\pm$0.13     & -2.16 (fix) & -0.036$\pm$0.019& 19.62$\pm$0.47 & 29.96/34  \\
 \hline
A8&39  & 0 &96.75$\pm$0.69&0.14$\pm$0.07& 15.22$\pm$0.46      &21.60$\pm$0.35      & -3.66$\pm$0.12     & -2.16 (fix) & -0.039$\pm$0.018 & 19.54$\pm$9.46 & 29.84/34  \\
\hline
T7& 87 & ff    &100.00$\pm$0.23 &0.08$\pm$0.09 & 16.07$\pm$0.43      &24.38$\pm$0.49      & -4.35$\pm$0.37     & -2.14 (fix) & -0.018$\pm$0.021 & 20.35$\pm$0.43 & 92.51/82  \\
\hline
T7& 87& 0     &99.94$\pm$0.25& 0.09$\pm$0.05 & 15.95$\pm$0.41      &24.26$\pm$0.50      & -4.25$\pm$0.36     & -2.14 (fix) &-0.021$\pm$0.012 &20.23$\pm$0.41 & 94.38/82  \\
\hline
A7 &40&ff    &95.76$\pm$0.16&0.00$\pm$0.04 & 15.94$\pm$0.17      &21.86$\pm$0.50      & -3.84$\pm$0.29     & -2.14 (fix) & 0.000$\pm$0.010 &20.22$\pm$0.17 & 21.01/35  \\
 \hline
A7&40 &0   &95.69$\pm$0.16& 0.00$\pm$0.12 & 15.89$\pm$0.17      &21.76$\pm$0.51      & -3.78$\pm$0.29     & -2.14 (fix) & 0.000$\pm$0.032 &19.46$\pm$0.17 & 20.62/35  \\
 \hline
  \hline
      &     &     &&   & &              &   &  &  &   &         \\
 
   \multicolumn{11}{c}{Condition IV) - $\rho = 0.14 $ and   $\mu_I = - 2.16 \GeV^{-2}$ \cite{LHC8TeV} or $-2.14  \GeV^{-2} $ \cite{LHC7TeV}
fixed  as in II) and III) }         \\
      \hline
      \hline
T8& 60& ff    &102.40$\pm$0.15& 0.14 (fix)& 15.27$\pm$0.39      &21.15$\pm$0.39      & -3.69$\pm$0.15     & -2.16 (fix) &   -0.038$\pm$0.002 &19.59$\pm$0.39 & 69.2/56  \\
\hline
T8& 60  & 0   &102.30$\pm$0.14& 0.14 (fix)& 15.23$\pm$0.07      &20.97$\pm$0.40      & -3.61$\pm$0.15     & -2.16 (fix) & -0.039$\pm$0.002 &19.55$\pm$0.07 & 63.14/56  \\
 \hline
A8&39 & ff  &96.82$\pm$0.11& 0.14 (fix) & 15.26$\pm$0.06      &21.65$\pm$0.24      & -3.69$\pm$0.12     & -2.16 (fix) &  -0.038$\pm$0.001 &20.03$\pm$0.14 & 29.97/35  \\
 \hline
A8&39 & 0  &96.78$\pm$0.11&0.14 (fix)& 15.24$\pm$0.06      &21.61$\pm$0.23      & -3.66$\pm$0.12     & -2.16 (fix) &  -0.038$\pm$0.001 &19.99$\pm$0.14 & 29.84/35  \\
\hline
T7& 87  & ff   &99.80$\pm$0.21 &0.14 (fix) & 15.71$\pm$0.14      &24.26$\pm$0.47      & -4.24$\pm$0.31     & -2.14 (fix) & -0.033$\pm$0.002 &19.99$\pm$0.14 & 95.08/83  \\
\hline
T7 & 87  & 0  &99.72$\pm$0.21& 0.14 (fix) & 15.67$\pm$0.14      &24.15$\pm$0.48      & -4.16$\pm$0.31     & -2.14 (fix) & -0.034$\pm$0.003 &19.95$\pm$0.14 & 94.98/83  \\
 \hline
A7  &40 &ff  &95.75$\pm$0.16&0.14 (fix) & 15.23$\pm$0.11      &21.86$\pm$0.44      & -3.99$\pm$0.22     & -2.14 (fix) & -0.035$\pm$0.002 &19.51$\pm$0.11& 27.33/36  \\
 \hline
A7  &40&0   &95.68$\pm$0.16&0.14 (fix) & 15.19$\pm$0.11      &21.77$\pm$0.44      & -3.94$\pm$0.22     & -2.14 (fix) & -0.036$\pm$0.002 &19.47$\pm$0.11 & 26.34/36  \\
\hline
\hline
\end{tabular}
 \caption{Results of fits for the four datasets. In Condition I) all (six)  parameters are free. 
In Conditions II, III and IV the parameters $\rho$ and  $\mu_I$ are restricted , as specified in the 
Table. In all cases comparison is made with results obtained with Coulomb-Nuclear interference phase 
$\phi$ put equal to zero ($\phi_0=0$) and calculated with electromagnetic proton form factor 
($\phi=\phi_{ff}$) as described in    Appendix A.
 }
\label{Table:All}
\end{center}
\end{table*}

The range of $-t$ covered in this  analysis corresponds to a forward region, with $|t|$ 
from $|t|_{\rm min}=0.00074$ GeV$^2$ 
for T8 to $|t|_{\rm max}=0.3709$ GeV$^2$ for T7. In this range the Coulomb effects play 
important role and the relative 
Coulomb phase is taken into  account. We compare results for the the relative phase $\phi=\phi_{ff}$, 
calculated in Appendix A  with proton form factor ($ff$) and the reference case of phase zero
 $\phi=\phi_{0}=0$. This alternative is examined to set reference values because of possible lack 
of understanding of  electromagnetic effects, as in the calculation of the phase and the possible 
influence   of the proton  radius at  high energies.

The statistical methods used in analyzes are performed with CERN-ROOT   software 
\cite{root}, accounting for statistical and systematics uncertainties. However, since the values 
of $\chi^2$ do not change much compared with the statistical uncertainties only, we understand 
that the statistical errors are enough for  the analyses with our amplitudes. 

We also study  the correlation between the parameters, which is an useful tool to control possible 
instabilities of the fits. The correlations between the parameters are defined as
\begin{eqnarray}
{\rm corr}(\alpha,\beta)=\frac{\langle \alpha \beta \rangle-\langle \alpha\rangle\langle\beta \rangle}{\sqrt{\sigma_{\alpha}^2\sigma_{\beta}^2}} ~ ,
\label{correlation}
\end{eqnarray}
where $\alpha$ and $\beta$ are any two parameters, the brakets are the expectation values of the 
fitted  parameters and $\sigma_\alpha$ and $\sigma_\beta $ the variances associated with  them. 
This correlation criterion  is known as Pearson coefficient and varies from -1 to 1, where -1 
is a complete anti-correlation,  meaning that if one parameter is increased the other one decreases, and 1 
is a complete correlation which means that if one parameter is increased the other one also increases.
 If the correlation coefficient is zero the parameters are said non-correlated, or independent. The  correlation factors help us to understand the relation between the  determination of the parameter and the 
range of $ |t|$  where this determination is performed. 
   
With the purpose of identifying generic or universal values for parameters, 
 for all measurements  we study four different conditions  in the fit: 
\begin{itemize} 
  \item  I) all six  parameters  are  free  ;  
  \item II) fixing $\rho$ at 0.14, as suggested by dispersion relations;  
  \item III) fixing $\mu_I$  from  the expected positions of   imaginary zero \cite{LHC8TeV,LHC7TeV} and dip in $d\sigma/dt$ ; 
  \item  IV) fixing simultaneously $\rho$ and $\mu_I$ at the above values. 
\end{itemize}

We present our analysis  for the four experiments  separately in the next subsections. Since T8  
has more precise  data in the very forward region and the experimental paper 
  has provided a detailed description of  the observed structure, we investigate it with more detail.  
For this purpose we  introduce a new diagram to represent the structure in the data at low $|t|$, 
plotting the ratio $T_R^2/T_I^2$ against $|t|$, 
with 
\begin{equation}
\frac{T_R^2}{T_I^2}=\frac{(\rho-\mu_R t)^2}{(1-\mu_I t)^2} ~  e^{(B_R-B_I)t} ~.
\label{ratio}
\end{equation} 
This ratio does not depend on the total cross section, and therefore   normalization 
uncertainties are cancelled, allowing identification of the zero   in the real amplitude.

The results of the analysis are presented  in Table \ref{Table:All}. The headings of the table indicate the 
  quantities determined in fits, namely the six parameters $\sigma$,$\rho$,$B_I$,$B_R$, $\mu_I$
and $\mu_R$. Other columns give the derived quantities $t_R$ and $B_I^{\rm eff}$, and the estimated 
$\chi^2$ values. The first three columns specify the measurement, with the number N of data points,
and  the phase option (either  the true 
phase $\phi_{ff}$ or the reference option of zero phase). 

\clearpage 

\subsection{T8}

\begin{figure*}[b]
  \includegraphics[width=8cm]{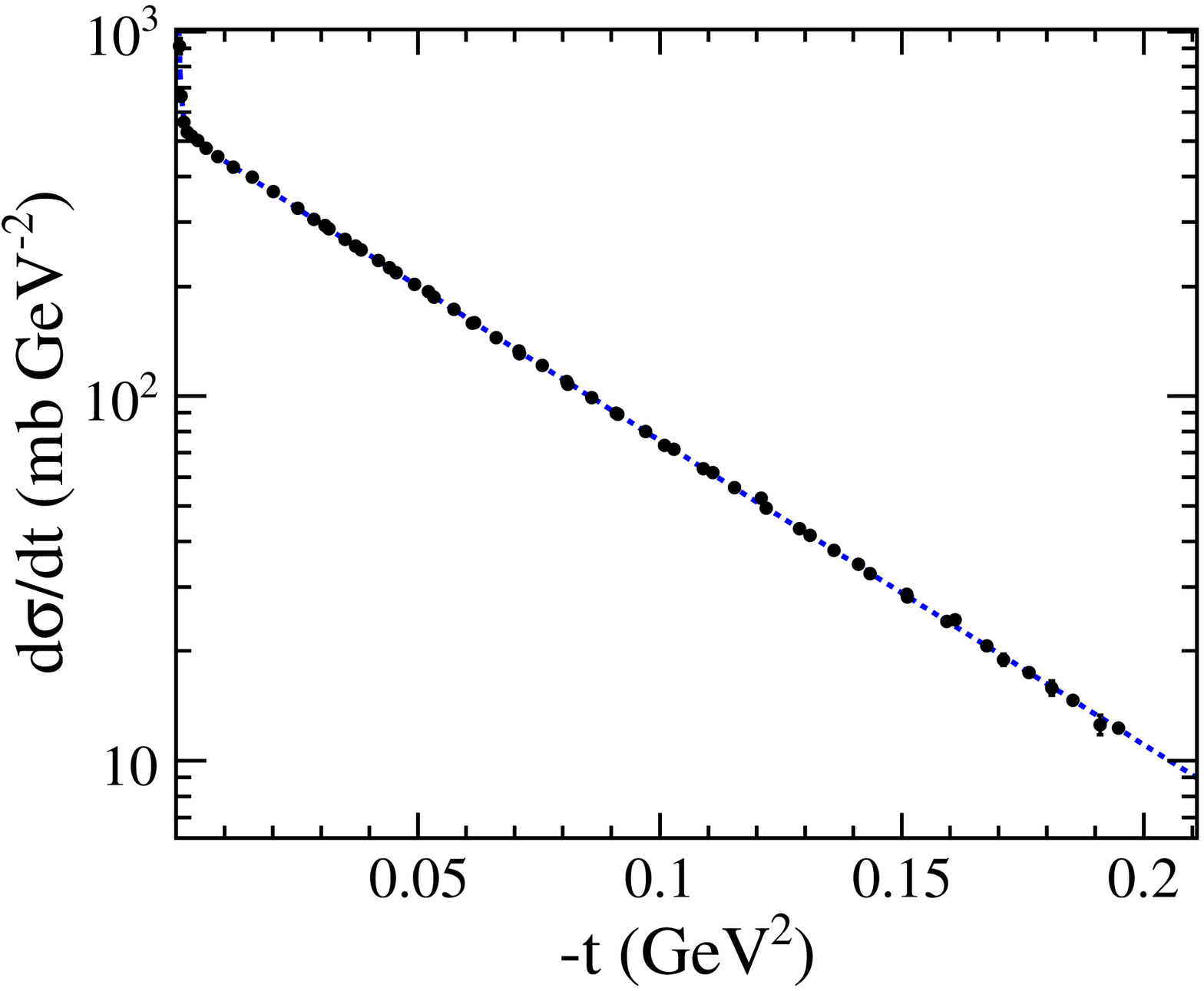}  
  \includegraphics[width=8cm]{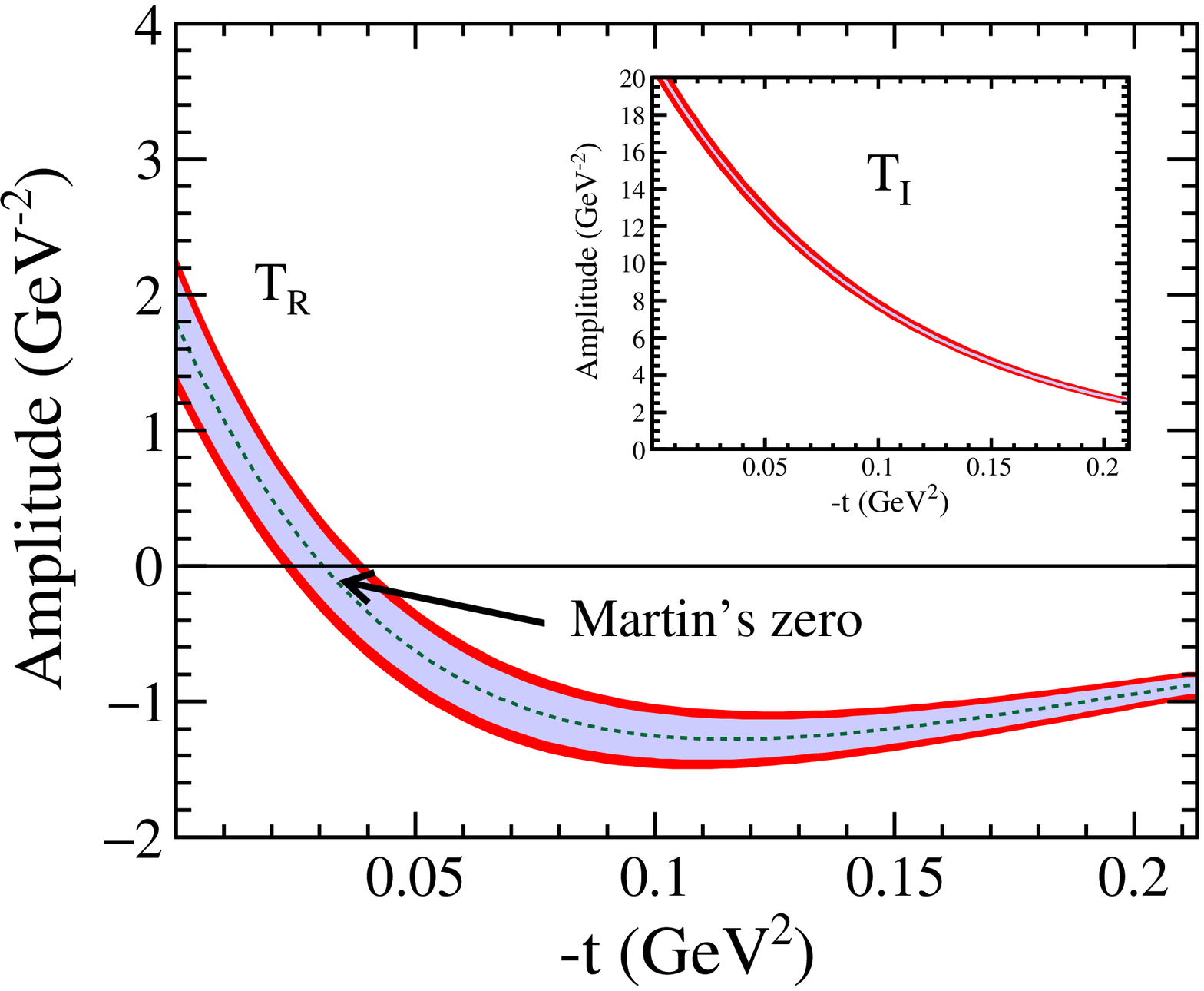}   
\caption{Result of the T8 data $d\sigma/dt$  fit (left)  
and the corresponding real and imaginary amplitudes (right). } 
 \label{dsigdt_amplitudes} 
\end{figure*} 
\begin{figure*}[b]
  \includegraphics[width=8cm]{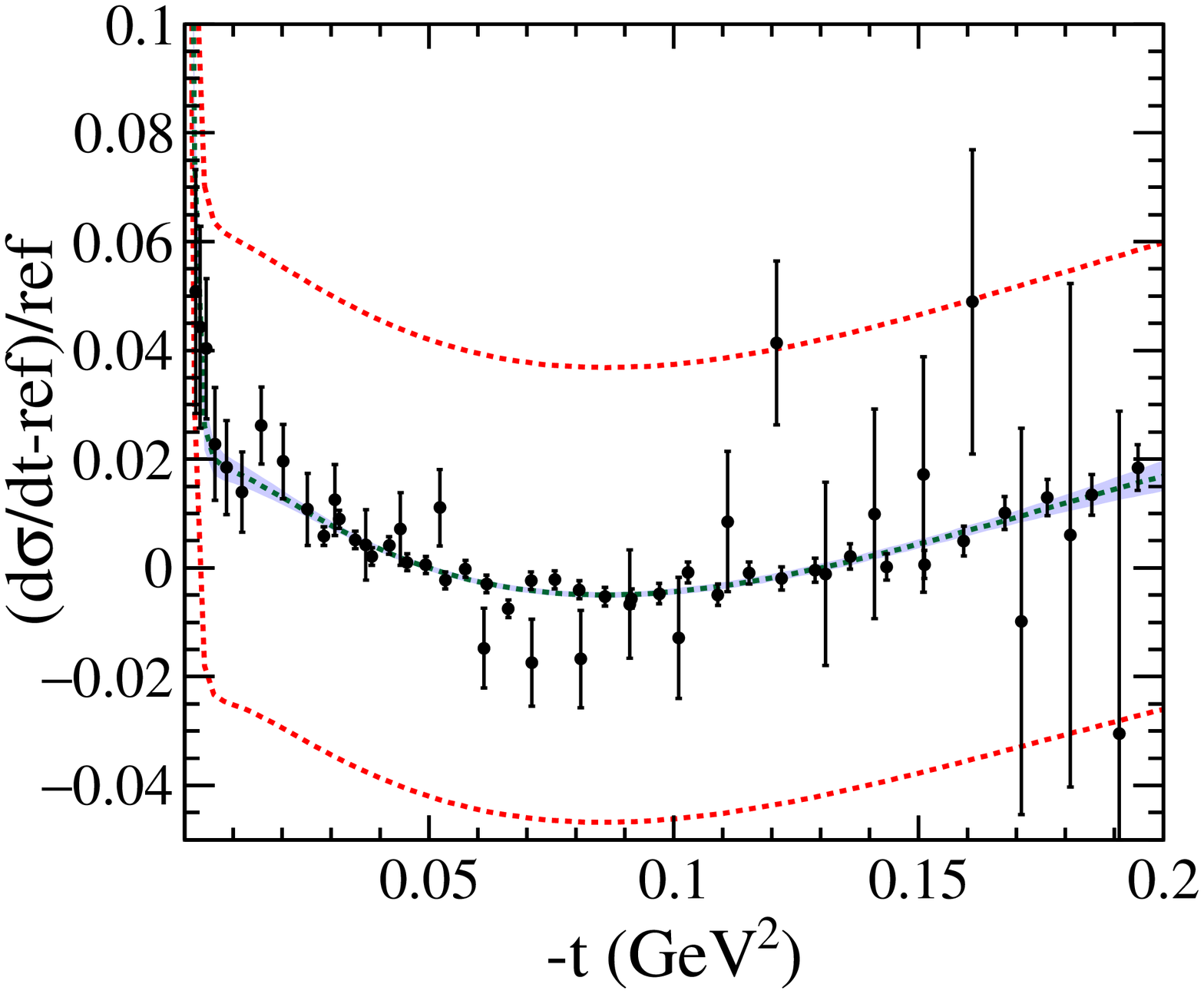} 
  \includegraphics[width=8.3cm]{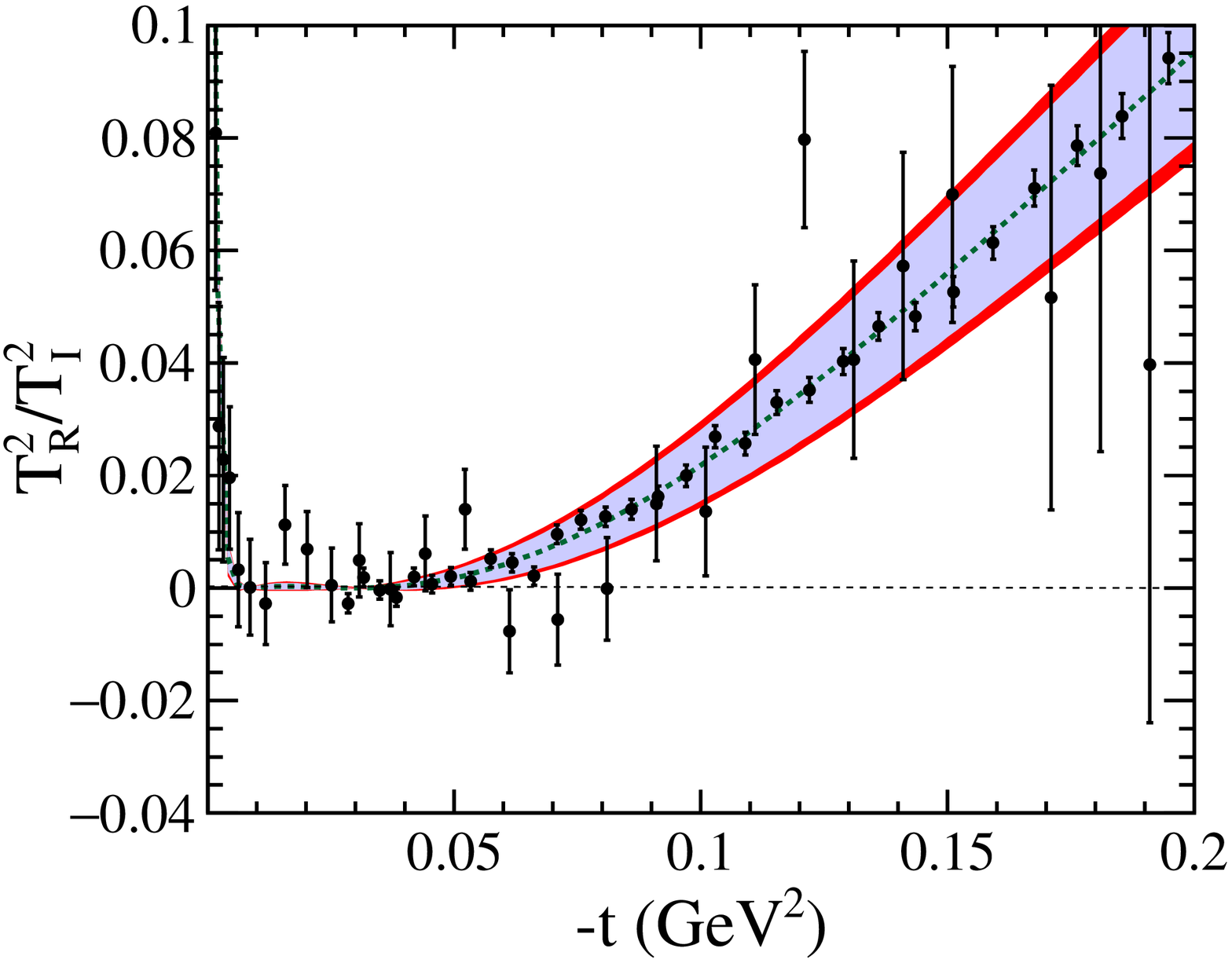}  
\caption{The left  plot  shows the non-exponential behaviour of the differential 
cross section for T8. The figure is obtained subtracting from the best fit  of the 
differential cross section a reference function which is $d\sigma/dt$ written 
 with a pure exponential form and dividing the subtraction by this reference function. 
The dashed lines show the normalization error band
in $d\sigma/dt$, that is quite large. The  plot in the RHS shows the ratio $T_R^2/T_I^2$ 
which exhibits information of a non-exponential behaviour with advantages compared with 
the first plot, since $\sigma$ is cancelled, and with it most of normalization
systematic error, and the second advantage is the clear indication of the structure of 
the real amplitude showing the position of its zero $|t|_R$.}
\label{displacement} \end{figure*} 
\begin{figure*}[b]
\includegraphics[width=5.5cm]{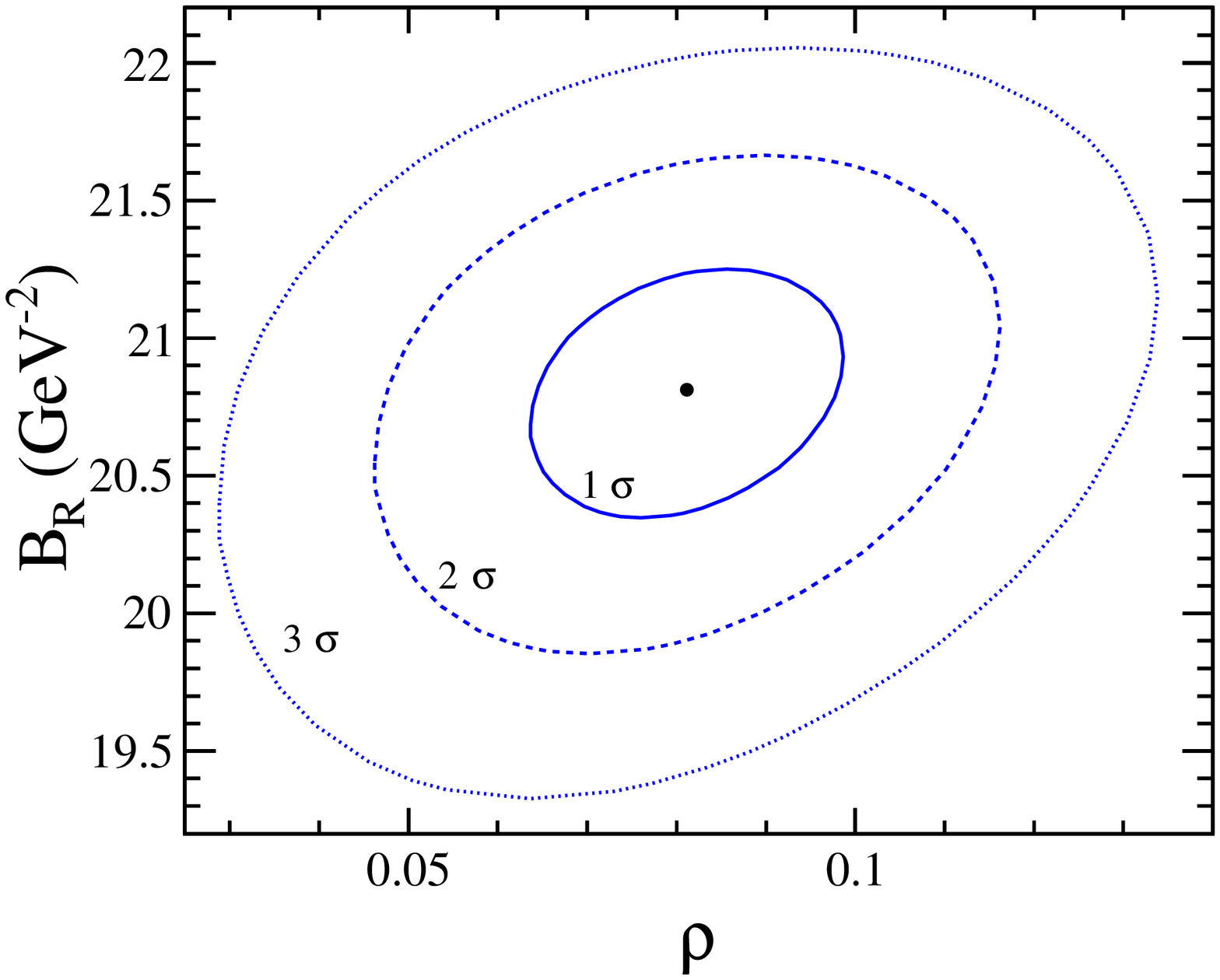}  
\includegraphics[width=5.5cm]{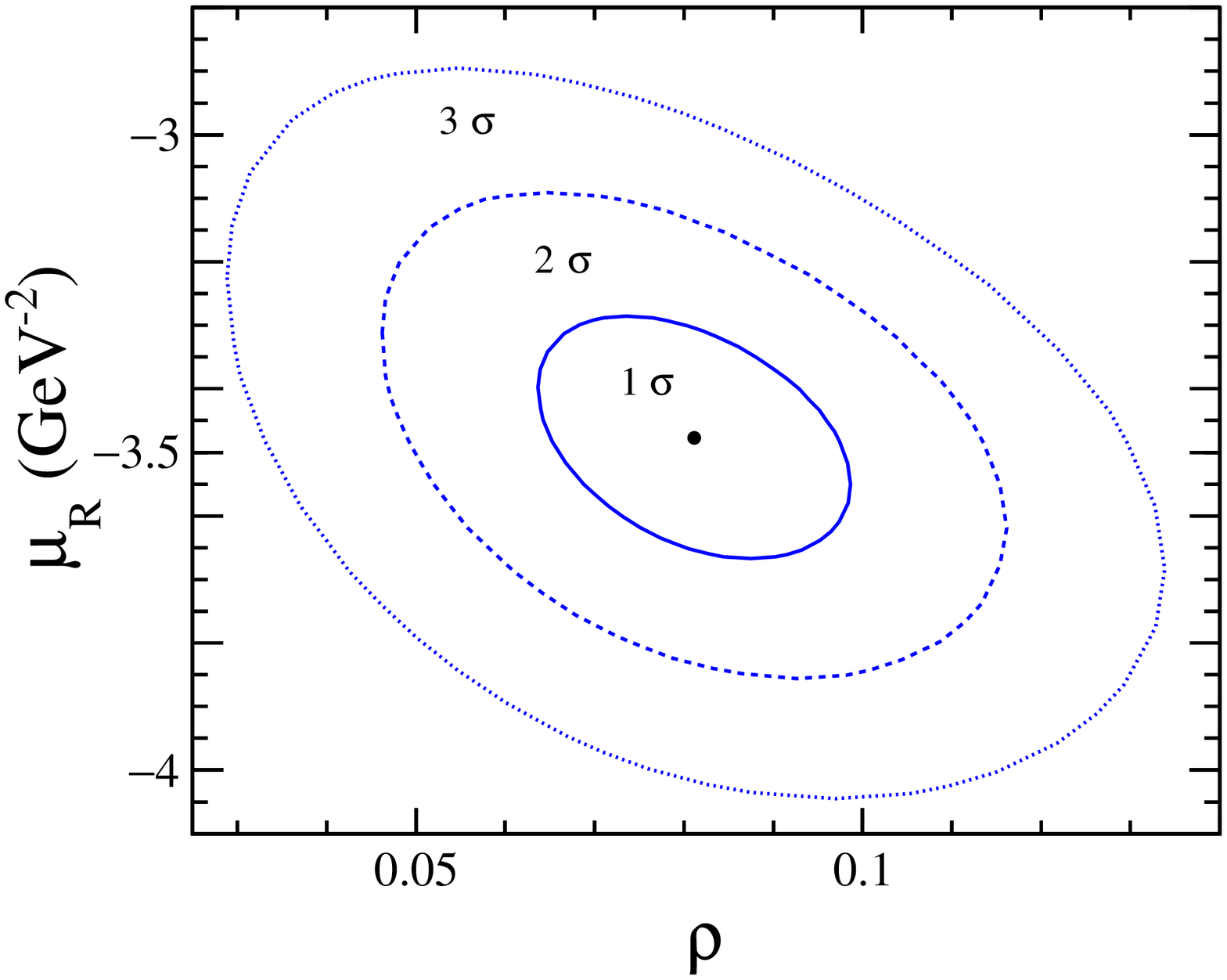}  
\includegraphics[width=5.5cm]{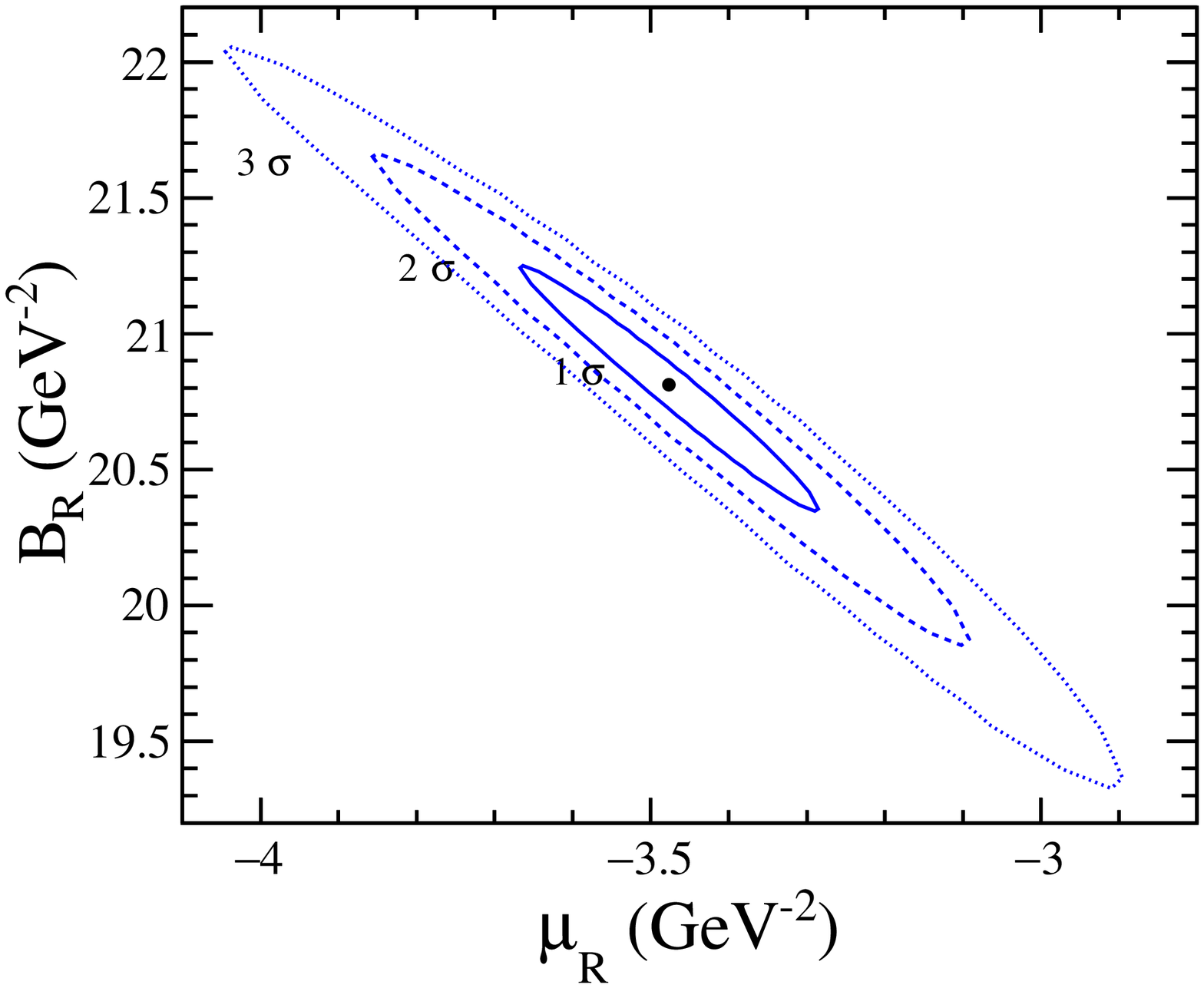}  
\caption{Correlation maps of the parameters of the real part for T8 measurements  under 
condition III). The flattening behaviour of the curves relating $(B_R, \rho )$ and 
$(\mu_R, \rho)$  show the weak  correlation between   parameters. The narrow areas in  
  $(B_R, \mu_R )$ indicate the strong correlation between these two 
parameters. }
\label{correlations} \end{figure*}

The T8 dataset contains measurements with two different optics. The first set of 
data (30 points) covers a very forward 
region, $0.000741 < -t <0.191  ~\GeV^2 $,  and the second set (30 points)
 starts a bit latter, $0.0285 < -t < 0.195 ~ \GeV^2 $, 
    overlapping partially with the first,  but  with  better statistical 
precision. They analyzed 
separately the first set, here called SET I,  with 30 points, and the combination of the 
first and the second measurements, with N=60 points, here called SET II. 
Analyzing   SET I  they obtained  $\rho=0.12\pm 0.03$. In the analysis with SET II, 
$\rho$ was not independently determined, but it was rather 
  fixed at the same value of SET I.  

Repeating this analysis in two steps, using our nuclear amplitudes and testing phase values  
$\phi=\phi_0=0$ and $\phi=\phi_{ff}$, we find that the determination of the $\rho$ parameter 
is critical, strongly influenced by points in the most forward region. Using Condition I), with all 
six parameters free, the results are the following. With SET I  we obtain  $\rho=0.106\pm 0.021$ 
for $\phi_{ff}$ case, which is a little bit bellow the expectation 0.14 from dispersion relations 
With the zero reference phase $\phi_0$ we obtain  $\rho=0.133\pm 0.021$.
With the complete  SET II  the $\rho$ values come out considerably smaller than in SET-I  
for both $\phi_0$ and $\phi_{ff}$ phases, deviating still more strongly from 0.14. 
These results show the difficulty of T8 for free independent determination of the 
parameter $\rho$.

SET II is used in the description that follows, and the resulting parameters are registered in  
Table \ref{Table:All}. 
 
Using the condition I) we obtain equivalent $\mu_R = -2.78 $ GeV$^{-2}$ for both $\phi_{ff}$ and 
$\Phi_{0}$ phases, but since the values of $\rho$ are different the position of the zero 
$t_{R}=\rho/\mu_R$  changes  from -0.032 $\pm$ 0.004 GeV$^{2}$ to -0.040 $\pm$ 0.004 GeV$^{2}$
for the $\phi_{ff}$ and $\phi_{0}$ phases respectively. We remark that precise data in the 
vicinity of the zero is important for the determination of $\rho$.  

In Fig.\ref{dsigdt_amplitudes} we present the fit  results of the T8 data 
with all parameters free, as in Condition I, and phase $\phi_{ff}$.  
The parametrization is able to describe the T8 measurements. 
  With plots showing local displacements, shown  in Fig. \ref{displacement}, 
we are able to exhibit details of the structure of the amplitudes.  
In the LHS we  plot   diagram similar to that presented in the paper of the 
experimental group \cite{T8},  showing the valley structure appearing when we subtract 
the simple exponential form 
            $$ ref ~= ~A ~ e^{Bt} $$  
from the best fit  solution for the differential cross section. 
The   structure appears neatly, but the band of normalization 
errors is very large, and the roles played by the  amplitudes are not clear. 
 In the right hand side of Fig. \ref{displacement} the quantity  $T_R^2/T_I^2$  
shows the turning point due to the zero of the real part of the amplitude, 
with a much narrower  band of 
the systematic errors due  to the cancellation of $\sigma$ 
dependence in the ratio.

The interplay of the magnitudes of the real and imaginary amplitudes influences the structure 
of $d\sigma/dt$. The deviation of a pure exponential behaviour is inherent to the sum 
of two independent squared quantities.
 
From the fits performed  in T8 we obtain from the correlation coefficients that 
$\rho$  is  weakly correlated with all  other parameters. 
This is the reason why when we fix its value according to Conditions II) and   IV) 
we obtain $\chi^2$   larger  than with Condition I): other parameters are not 
able to compensate the change in $\rho$.

In both conditions I) and II) the parameter of the imaginary part $\mu_I$ is 
very small, which means that the t-range in the data is not far 
enough to feel  the zero of the imaginary amplitude.  
On the other hand, fixing   $\mu_I$ in Condition III), $\rho$ is seen as strongly 
dependent on the phase. 

  It is important that, although the parameter $\mu_I$ has weak Pearson coefficients, 
the statistical error associated with this parameter is large, which means that it 
cannot be well determined with Condition I) or II). The $\chi^2$ 
values do not  change considerably from Condition I) to Condition III), but the 
parameters $B_I$, $B_R$ and $\mu_R$ change. The presence of non-zero 
negative  $\mu_I$  obliges the imaginary amplitude to point towards zero.  
The real amplitude must compensate the decrease of the imaginary part, reducing 
the magnitude of the real slope, and  the value  of $\mu_R$ is also affected. 
The imaginary slope compensates the increase in the 
magnitude of $\mu_I$  introduced in III), thus preserving the value of $B_I^{eff}$. 

    Fig. \ref{correlations} shows the correlation maps between pairs of parameters. 
These figures correspond to Pearson coefficients ${\rm corr} (B_R,\rho)=0.26$, 
 ${\rm corr} (\mu_R,\rho)=-0.39$ and ${\rm corr} (B_R, \mu_R)=-0.98$
showing weak correlations for the first two cases, and a strong anti-correlation
for the latter. 
 The lines  represent the  allowed regions at different standard deviations.  
Since $\rho$ is weakly correlated with the parameters of the real 
part according to Condition III), in Condition IV)  we expect small 
deviation in $B_R$ and $\mu_R$. In this Condition we obtain very similar 
results for the free parameters for both choices of  phase. 
 
\subsection {A8}

\begin{figure}[b]
\includegraphics[width=8cm]{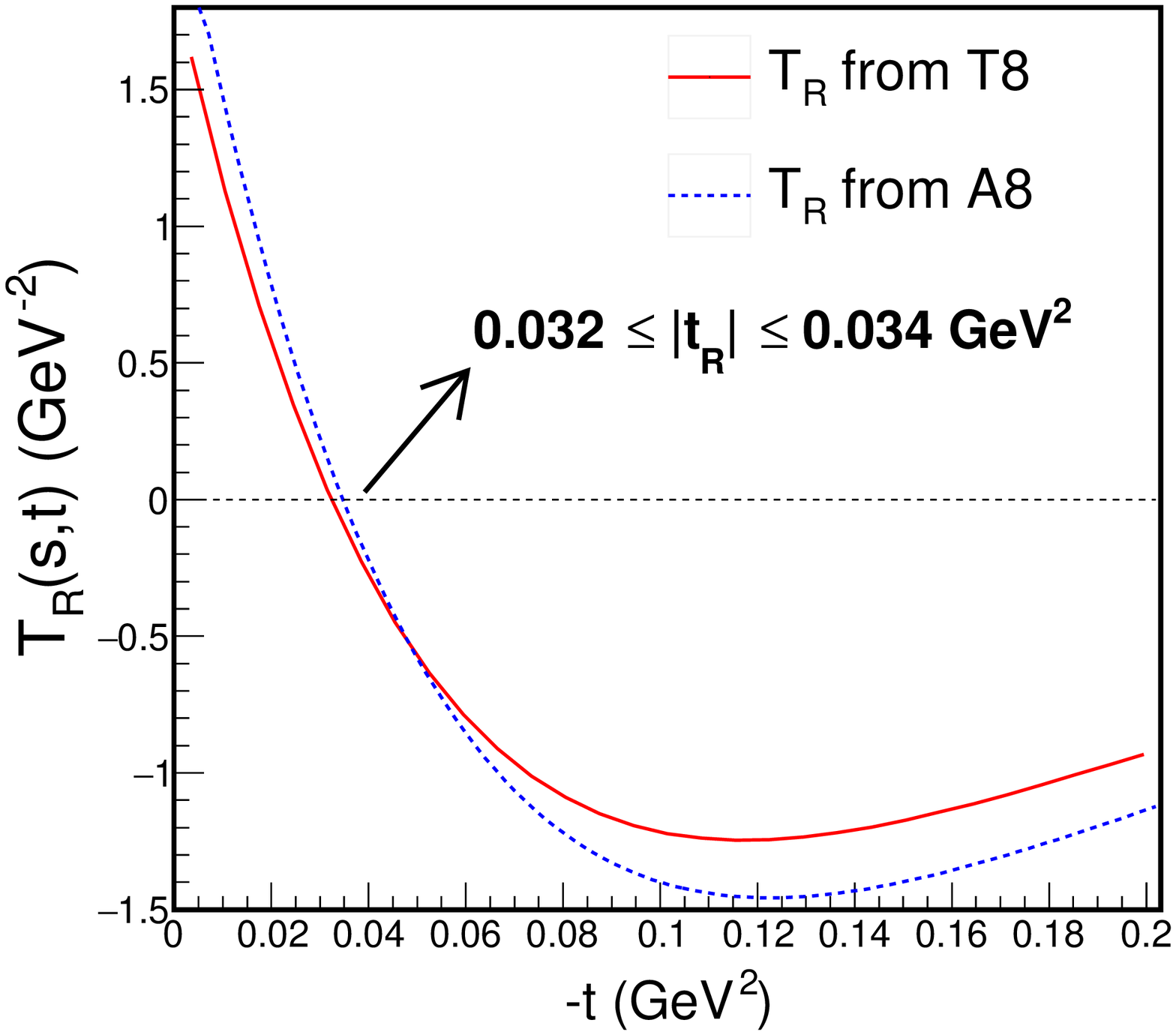}  
\caption{Comparison of the real amplitudes in T8 (solid lines)  and A8 
(dashed lines) measurements.     These amplitudes represent Condition I), 
with all parameters free and phase $\phi_{ff}$. 
It is remarkable the (almost) coincidence of the zeros in T8 and A8. }
\label{real_ampl} \end{figure}

 In A8 \cite{A8} the  Atlas Collaboration measured   39 points in the region 
$0.0105<-t < 0.3635 ~(\GeV^2) $. 
With  Condition I) we obtain central $\rho$ 
values compatible with dispersion relations (namely 0.14), but with large error bars 
of 70$\%$. Showing insensitivity, the values of $\mu_R$ are similar   for  
$\phi_{ff}$ and $\phi_{0}$ phases.       
   Although the $\mu_R$ values in A8 and T8 are different by  20\%, 
the positions of the real zero differs  about 6\%. The real amplitudes for 
T8 and A8 are shown in Fig. \ref{real_ampl} and we can clearly note that 
the position of the Martin's zero is in agreement between the measurements.

Important differences between T8 and A8 are in the  parameters of the imaginary part. 
Thus $\mu_I=-1.65 \GeV^{-2}$ for A8,  while for T8 it is compatible with zero due to 
the short $|t|$ range of the data. 
 
The values of $\sigma$ in A8 and T8  differ by $\sim 15\%$. This difference 
may be due relative normalization, and we may ask whether a unification of  
$d\sigma/dt$ through a constant factor could  unify the solutions for $\sigma$, while 
leading  also $\rho$ and $\mu_R$  to common  values.  
 
    Since  the fits for A8 show strong anti-correlation factor 
between $\sigma$ and $\rho$, and $\sigma$  is a stable parameter,  we use Condition II) 
to fix $\rho$ at 0.14 and  obtain $\chi^2$ very similar to the result obtained in Condition I)  
with free $\rho$, which is natural since in I), $\rho$ in not very different from 0.14. 
 Under Condition II), with fixed $\rho$,  the   results for T8 and A8 give compatible 
$\mu_R$ values.

In  A8, for both I) and II),  the $\mu_I$ parameter predicts $|t_I|=0.59 \GeV^{2}$ 
which is  far on the right of the region where the dip in $d\sigma/dt$ is expected to occur. By fixing $\mu_I$ in Condition III), $B_I$ for A8 agrees with the value found 
in T8, as expected, since values of $B_I^{eff}$ are in agreement for all Conditions.
  
 The use of Condition IV) does not change   considerably the parameters 
of A8 when   compared with Condition III), because the central values of 
$\rho$ in  III) are  close to 0.14 .

\clearpage
 
\subsection{T7 }

 At $\sqrt{s}$=7 TeV the  Totem Collaboration measured elastic p-p cross
sections with two sets \cite{T7}, the first in the range ($0.005149 < -t< 0.3709$ GeV$^{2}$ ) 
with 87 points, and the second   in the range $0.377 < -t< 2.443$ GeV$^{2}$ 
with 78 points.  Using our expressions for the forward amplitudes, we analyze   the forward set, obtaining the results shown in Table \ref{Table:All}. With all parameters free, 
for both $\phi_{ff}$ and $\phi_0$ phases the $\rho$ values are  less than 0.14.   
 The   values for $\mu_I$ are similar to those in A8. The total cross section is higher 
but compatible with the original  paper \cite{T7}, as given in Table \ref{datasets}.  
 
 The    correlation   factor ${\rm corr} (\sigma,\rho)=-0.95$  obtained in Condition I),  
shows a strong anti-correlation. Since $\sigma$ is a stable parameter and the central $\rho$ 
value is larger than the expected  0.14,   Condition II)   fixing  $\rho$, leads 
for all other parameters and for $\chi^2$    results similar to Condition I).  

  With Condition III)  
  we  observe that the central $\rho$ values are smaller than 
 obtained with I), and the statistical errors are about 
55$\%$ ($\phi_0$) and 88$\%$ ($\phi_{ff}$) of the central value. 

\bigskip
{\bf   Analysis of an  Extended Set } 
\bigskip

\begin{table*}[t]
\begin{center}
 \vspace{0.5cm}
\begin{tabular}{c c  c c c c c c c c c c}
\hline 
\hline 
 & N & $\phi$  &$\sigma$ & $\rho$        &$ B_{\rm I}  $      &$ B_{\rm R}  $      & $~\mu_{\rm R}  $     & $ \mu_{\rm I} $ & $t_R$ & $B_I^{\rm eff}$     & $\chi^2/$ndf\\
  &   &  &  (mb)       &               &$(\GeV^{-2})$ &$(\GeV^{-2})$ &$(\GeV^{-2})$ & $ (\GeV^{-2})$& $ (\GeV^{2})$ & $ (\GeV^{-2})$ &         \\
\hline
\hline
           &     &   & & & &            &   &  &  &   &         \\
   \multicolumn{11}{c}{Condition I) - all six parameters free}         \\
      \hline
      \hline
T7 & 87+17 & ff  &99.33$\pm$0.49 &0.16$\pm$0.05 & 15.36$\pm$0.24      &22.63$\pm$0.38      & -3.54$\pm$0.22     & -2.15$\pm$0.02& -0.045$\pm$0.014 & 19.66$\pm$0.24 & 203.1/98  \\
\hline
 T7 & 87+17& 0    &99.18$\pm$0.52& 0.17$\pm$0.05 & 15.30$\pm$0.24      &22.51$\pm$0.39      & -3.47$\pm$0.21     & -2.14$\pm$0.02& -0.048$\pm$0.014 & 19.58$\pm$0.24 & 202.1/98  \\
\hline
\hline
            &     &   & & & &            &   &  &  &   &         \\
   \multicolumn{11}{c}{Condition II) - $\rho = 0.14$ fixed by dispersion relations}         \\
      \hline
\hline
T7 & 87+17& ff     &99.49$\pm$0.18 &0.14 (fix) & 15.46$\pm$0.08      &22.73$\pm$0.30      & -3.57$\pm$0.22     & -2.15$\pm$0.02 & -0.039$\pm$0.002 & 19.76$\pm$0.09& 203.3/99  \\
\hline
T7& 87+17 & 0     & 99.42$\pm$0.17& 0.14 (fix) & 15.43$\pm$0.08      &22.65$\pm$0.30      & -3.52$\pm$0.22     & -2.15$\pm$0.01 & -0.040$\pm$0.002& 19.73$\pm$0.08& 202.4/99  \\
\hline
\hline
      &     &     &    &    &   &       &   &  &  &   &         \\
 
   \multicolumn{11}{c}{Condition IV) - Fixed values $\rho=0.14$ and $\mu_I=-2.14 \GeV^{-2} $  \cite{LHC7TeV}  }         \\
      \hline
      \hline
T7 & 87+17 & ff &99.44$\pm$0.14 &0.14 (fix) & 15.44$\pm$0.07      &22.62$\pm$0.19      & -3.49$\pm$0.13     & -2.14 (fix)& -0.040$\pm$0.002 & 19.72$\pm$0.07 & 203.5/100  \\
\hline
 T7& 87+17 & 0  &99.39$\pm$0.14& 0.14 (fix) & 15.43$\pm$0.07      &22.59$\pm$0.19      & -3.47$\pm$0.13     & -2.14 (fix)& -0.040$\pm$0.002 & 19.71$\pm$0.07 & 202.5/100  \\
\hline
\hline
\end{tabular}
 \caption{Fits for   the extended T7 data (N=104 points  in the range $0.00515 < -t<  0.52  ~\GeV^2$) according to Conditions I), II) and IV). }
\label{Table:T7extended}
\end{center}
\end{table*}

 The special availability in T7 of data beyond the $|t| \approx 0.4 \GeV^2$ limit 
of the measurements in Table \ref{datasets}, may be used to study a range where the 
$\mu_I$ parameter becomes more effective,  pointing towards the zero of the imaginary 
amplitude, and being determined with higher accuracy. We thus add to the forward set 
the first 17 points of the second dataset, reaching $|t_{\rm max}|=0.52 \GeV^2$. The 
results of the analysis with the combined set of 104 points are presented in 
Table \ref{Table:T7extended}.

\begin{figure}[b]
\includegraphics[width=8cm]{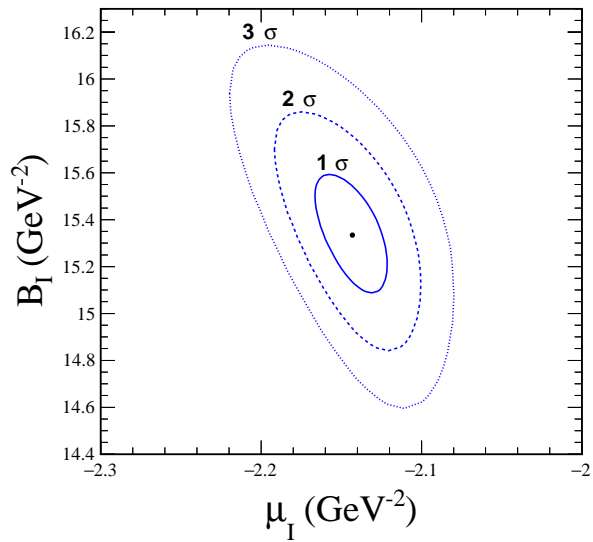}  
\caption{  Correlation map between the parameters $B_I$ and $\mu_I$ for the extended T7 set (104 points) studied under Condition I).}
\label{cor_mui_bi} \end{figure}

   The study of the extended set with Condition I) leads to  the correlation map for the 
quantities $B_I$ and $\mu_I$  shown in Fig. \ref{cor_mui_bi}.   
The Pearson coefficient depends on the range on $t$ where the fit is made and of course 
the fit conditions used, and for this extended range it shows a slight  anti-correlation between $\mu_I$ and $B_I$.  

\begin{figure}[b]
\includegraphics[width=8cm]{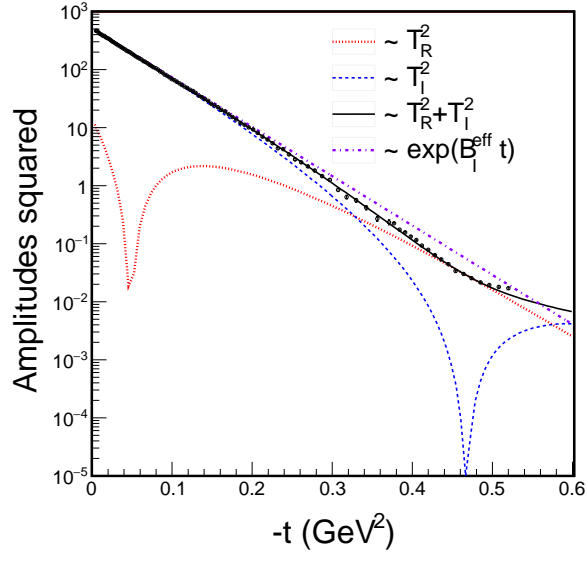}  
\caption{  Comparison of the squared  magnitudes of real, imaginary and simplified  single exponential amplitudes for the extended T7 dataset, with 104 experimental points (full circles).
The  real amplitude has a zero near 0.04$\GeV^2$ and its magnitude  competes
 with the imaginary amplitude from  $|t|=0.3$ GeV$^2$. The imaginary amplitude  deviates from simple exponential form near   $|t|\sim 0.2$ GeV$^2$ and passes through zero at $|t|=0.47$ GeV$^2$.}
\label{BI_eff_7_TeV} \end{figure}

In Fig. \ref{BI_eff_7_TeV} we compare the squared magnitudes of the real and imaginary 
amplitudes, and also the simplified single exponential amplitude assuming the effective 
slope $B_I^{\rm eff}$, for the extended T7 with Condition I). We observe that the imaginary 
part starts to deviate from the simple exponential near $|t|\sim 0.2$ GeV$^2$, and at 
$|t|\sim 0.47$ GeV$^2$ it passes through a zero. 
Beyond this range the real amplitude would be modified to incorporate other terms 
(say a quadratic $\xi_R t^2$) that may play important role in the construction of the dip structure. 
  
 As with the   forward set, here  the $\rho$ values come out larger than 0.14, but 
within error bars. Fixing $\rho=0.14$ in   Condition II), the changes in parameters and in $\chi^2$ are very small.

 It is very important that the parameter $\mu_I$ is here  determined with more precision and   is compatible with the value fixed to establish Condition III)  
in Table \ref{Table:All}.  
   
The use of condition IV) for this extended set, shown in 
Table \ref{Table:T7extended}, 
leads to a decrease in the magnitudes of $B_R$ and $\mu_R$ when we compare with the results of the forward set under the same condition. Since $\rho$ is fixed, the decrease in $\mu_R$ implies in an increase in the magnitude of $|t_R|$.

\clearpage

\subsection{A7}

 In A7  the  Atlas Collaboration measured \cite{A7} 40 points in the region 
$0.0062<-t < 0.3636 ~(\GeV^2)$.  This experiment is challenging, since the 
experimental authors recognize that $\rho$ cannot be determined from the 
data with usual forms of amplitudes. The natural
result  for $\rho$ as a free parameter is a negative quantity, and   
we run Condition I) imposing  a lower bound 0 for $\rho$, and of course 
the expected value is zero,    with large error bars. In this 
Condition the total cross section for $\phi_{ff}$ phase  is  $\sigma = 95.77$ mb, 
which is well  bellow, while the parameters $B_I$, $B_R$, $\mu_R$ and $\mu_I$ 
are compatible with  T7. In these measurements the correlation factors between 
$\rho$ and all other parameters are very small. Thus with Condition II) 
we expect an obvious  increase of $\chi^2$, but the parameters 
$B_I$ , $B_R$ and $\mu_R$ are compatible with T7. 

 With Condition I),   $\mu_I$ is compatible with zero, with large error. 
With Condition III), $\chi^2$ is not much changed, as well as the 
total cross section. The $\rho$ value is still undetermined but $B_I$, $B_R$ 
and $\mu_R$ are compatible with the other  experiments.  Thus, in spite  of the smaller 
value of total cross section and instabilities in the determination of $\rho$, 
Condition IV) shows similar $t$ dependence in A7 with respect to the other datasets, 
with  similar behaviour of the amplitudes. 


\section{ Summary of the Analysis }

The LHC measurements at $\sqrt{s}$  = 7 and 8 TeV shown in
Table I are analyzed under the assumption of the analytical 
forms for the scattering amplitudes given in Eqs. 
in Eqs. (\ref{imag_TI}) and (\ref{real_TR}).
  These forms are considered  as simple as possible
under theoretical conditions to describe  
the scattering amplitudes in the  forward region. 
The analysis aims at the determination of six intervening parameters,
three for the imaginary part ($\sigma$,$B_I$,$\mu_I$) and three 
($\rho$, $B_R$, $\mu_R$) for the real part, with 
  expected smooth energy dependence. 
 Only the T8 measurements cover very small $|t|$ range, so that 
the analysis is made with  non-homogeneous inputs, and   four specific 
Conditions, named I), II), III), IV) in Table \ref{Table:All},  are studied 
separately.

The Coulomb-Nuclear interference is a crucial ingredient, and its phase 
$\phi_{ff}$ is treated  compatibly with the forms of the amplitudes, 
as presented in Appendix A. 
 In order to have a reference (although not realistic) we also give 
results of fits with phase $\phi_0$ put equal to zero. 
 
As expected, the direct results of the fits with all parameters kept free are rather 
dispersive in some aspects, as shown in the sub-table with title Condition I).
The fitted values of $\rho$ do not agree among the measurements within 
the statistical uncertainties, but once the normalization uncertainty is 
considered, the values are in agreement.  
 We observe that the other parameters of the real part ($B_R$
and $\mu_R$), related to the shape of the amplitudes but not so strongly to its 
normalization, appear  more regular.

In Condition II)  $\rho$ is fixed at a reference value 0.14  suggested by dispersion 
relations for the 7-8 TeV range. Compared to I) there is a loss in $\chi^2$
in the cases with free $\rho$ far from  0.14, namely T8 and A7, but not in the 
other cases, where we  observe only rather slight adaptation in the other real 
parameters: $B_R$ and $\mu_R$ appear as regular quantities. Also $B_I$ remains the same 
(except in A7), and it is particularly important to remark  that  the effective slope 
  $B_I^{\rm eff}=B_I-2 \mu_I $  in Eq. (\ref{BI_eff}),  
that compensates for the influence of $\mu_I$ on the $|t|$ dependence of 
the imaginary part, appears as a very regular common  quantity. 
In spite of the differences, this point we may say that we 
are lead accept the value $\rho \approx 0.14$ for all measurements. On the other hand,  
    $\mu_I$ remains not regular among the experiments in this 
Condition II).  Since $\mu_I$ is responsible for the 
presence of a zero ($t_I=1/\mu_I$) in the imaginary amplitude (that occurs near the dip
in $d\sigma/dt$ at about 0.4 $\GeV^2$), it is natural that T8 
(limited to 0.2  $\GeV^2$) is not sensitive to $\mu_I$, and puts it at zero
 in Conditions I) and II).

 It is interesting to observe the effects of correlations. For instance, 
 in  T8 and A7 ,  $\rho$ is weakly correlated with the other parameters. 
Then,  fixing $\rho$ in Condition II) worsens strongly $\chi^2$ for these 
datasets, more  than in sets A8 and T7 where the strong correlation between 
$\rho$ and the other parameters  absorbs the effects of the fixing condition. For  T8 in 
$\phi_{ff}$  case the worsening in $\chi^2$ is more dramatic,  
corroborating  our concern about proton form-factor  and Coulomb phase.

With Condition III) we  fix $\mu_I$ according with the expected 
positions of the imaginary zero and dip in $d\sigma/dt$
\cite{LHC8TeV,LHC7TeV}, and let $\rho$ free.  This is successful, 
as $\rho$ results about the  same or improved with respect to 
Condition I) (all parameters free), 
except for A7, as expected. 
The   parameters $B_R$ and $\mu_R$ of the real part remain the same, 
and it is remarkable that $B_I$ changes, becoming  very regular, 
absorbing the influence of $\mu_I$ (now fixed)  and keeping the constant 
and regular effective $|t|$ dependence of the imaginary amplitude, represented 
by $B_I^{\rm eff}$.

The $\mu_I$ parameter determines a zero at $t_I=1/\mu_I$ in the imaginary amplitude 
and is   related   with the position of the  dip in the differential cross section. 
A precise determination of $\mu_I$  depends on the existence of data in a region 
near the dip. 
 Thus in the  T8 dataset, without points for small $|t|$, the central values 
of $\mu_I$ are near zero in I) and II), while in A8, T7 and A7 the  
values   have larger magnitudes.
We study this question analyzing the T7 experiment with inclusion of a 
second set of points \cite{T7}. Forming a larger dataset in the range 
$0.005149-0.52 ~\GeV^2$, the best solution with fixed $\rho=0.14$  
gives  $\mu_I=-2.15 ~\GeV^{-2} $  shown in Table \ref{Table:T7extended},
in good agreement with the   prediction  \cite{LHC8TeV}.  
We are thus lead to Conditions III) and IV) that fix $\mu_I$.

Fixing both $\rho$ and $\mu_I$  at their {\it expected}  values
in  Condition IV) we obtain good modeling 
for all measurements, except for the total cross 
sections, that separate Atlas from Totem. 
  It is particularly meaningful that 
the position of the zero of the real amplitude is nearly the same for all  cases.

 The   deviation from pure exponential form  in the differential cross section   
is interpreted  as due to the shape difference between the real and imaginary amplitudes. 
 The T8 experiment presents very precise data at low $|t|$,  showing a valley
behaviour in the differential cross section, while the  T7, A7 and A8  data also indicate a structure  
for the data at low $|t|$,  but with large uncertainties. The shape  is constructed 
when the real amplitude crosses zero, passing  to negative values. 
After the zero, the action of the real slope pushes this negative value back to zero, and the   structure is formed. 

This mechanism suggests that the determination of $\rho$ depends not only on the 
extrapolation to the limit at $|t|=0$, but   also 
depends on the  form  of the real  amplitude around its zero. 
  A precise extraction of $\rho$ 
depends both on the specific analytical model used and on the data in the whole range
  $0<-t<0.2$ GeV$^2$ where the valley occurs.  
The position of the real zero is obtained from the equation $t_{R}=\rho/\mu_{R}$ 
(values are given in Table \ref{Table:All}). 

Important  quantities are the derivatives of the amplitudes and their effective slopes  
  determined in each measurement. 
The exponent $B_I$ written in Eq. (\ref{imag_TI}) 
is not the logarithmic derivative of  the imaginary amplitude, that is given by 
Eq.(\ref{BI_eff}). The  effective slope at small $|t|$ 
   $$   B_{I}^{\rm eff}\simeq B_I-2\mu_I   $$  
is seen  in the approximation of  the linear factor   
put   in exponential form $1-\mu_I t \approx \exp(-\mu_I t)$. Thus the  
determinations of $\mu_I$ and $B_I$ depend on data in the very forward region, and 
also in the region near the dip. Table \ref{Table:All} shows the interesting 
regularity of the quantity $B_I^{\rm eff}$,  as opposed to $B_I$.

The {\it average  slope} $B$   measured  in the differential cross section is
given in Eq.(\ref{slope_dsigmadt}). In our analysis $B$ appears  stable, with value  
  $B\simeq 20.5$ GeV$^{-2}$. Comparing our result with the 
  values in Table \ref{datasets} we see deviations of about 1 GeV$^{-2}$, and 
we thus remark that the measured {\it average  slope} $B$ depends on parameters 
$\mu_I$ and $\mu_R$ that are influenced by data in the  large  $|t|$ region. 

\begin{figure*}[b]
\includegraphics[width=8cm]{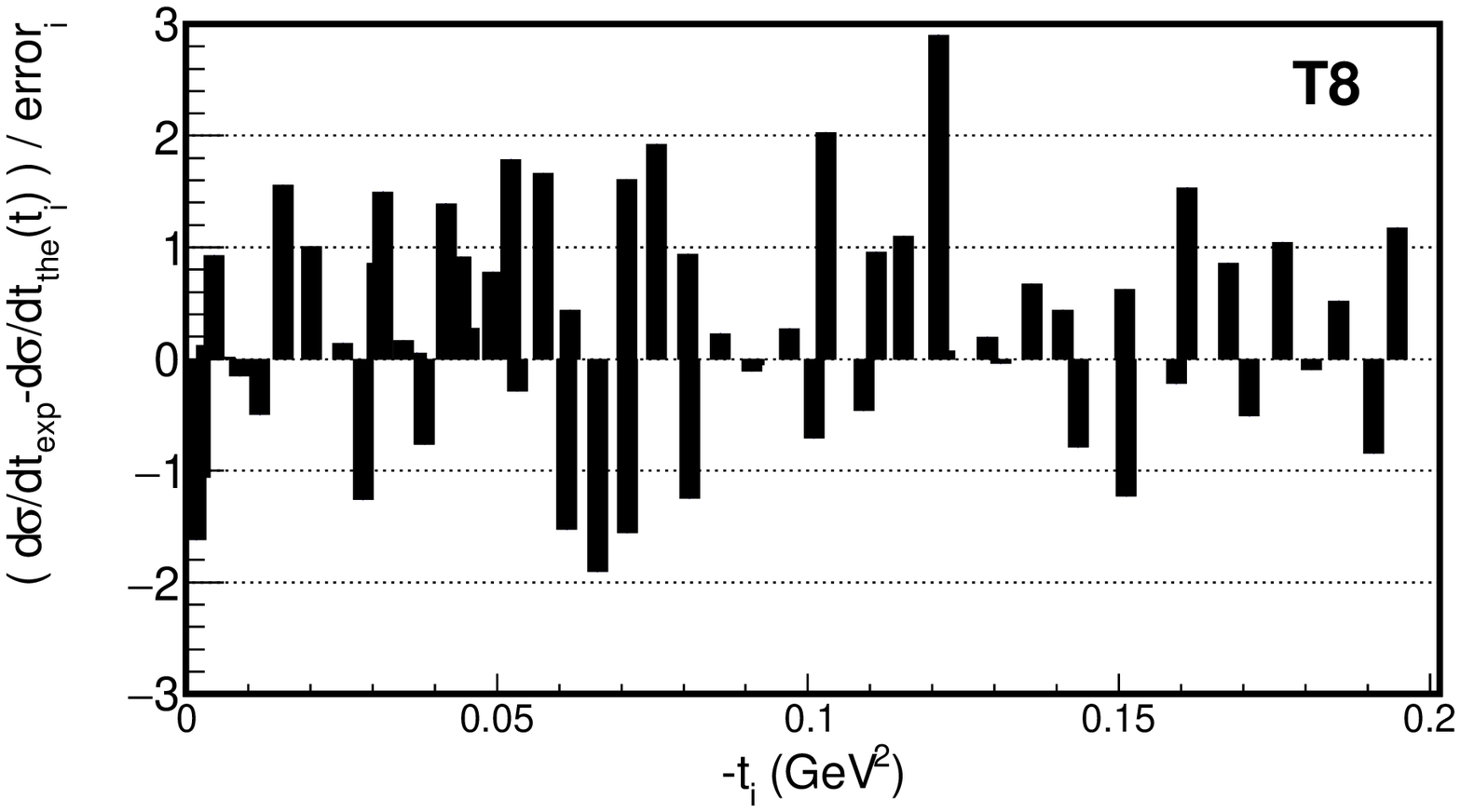}  
\includegraphics[width=8cm]{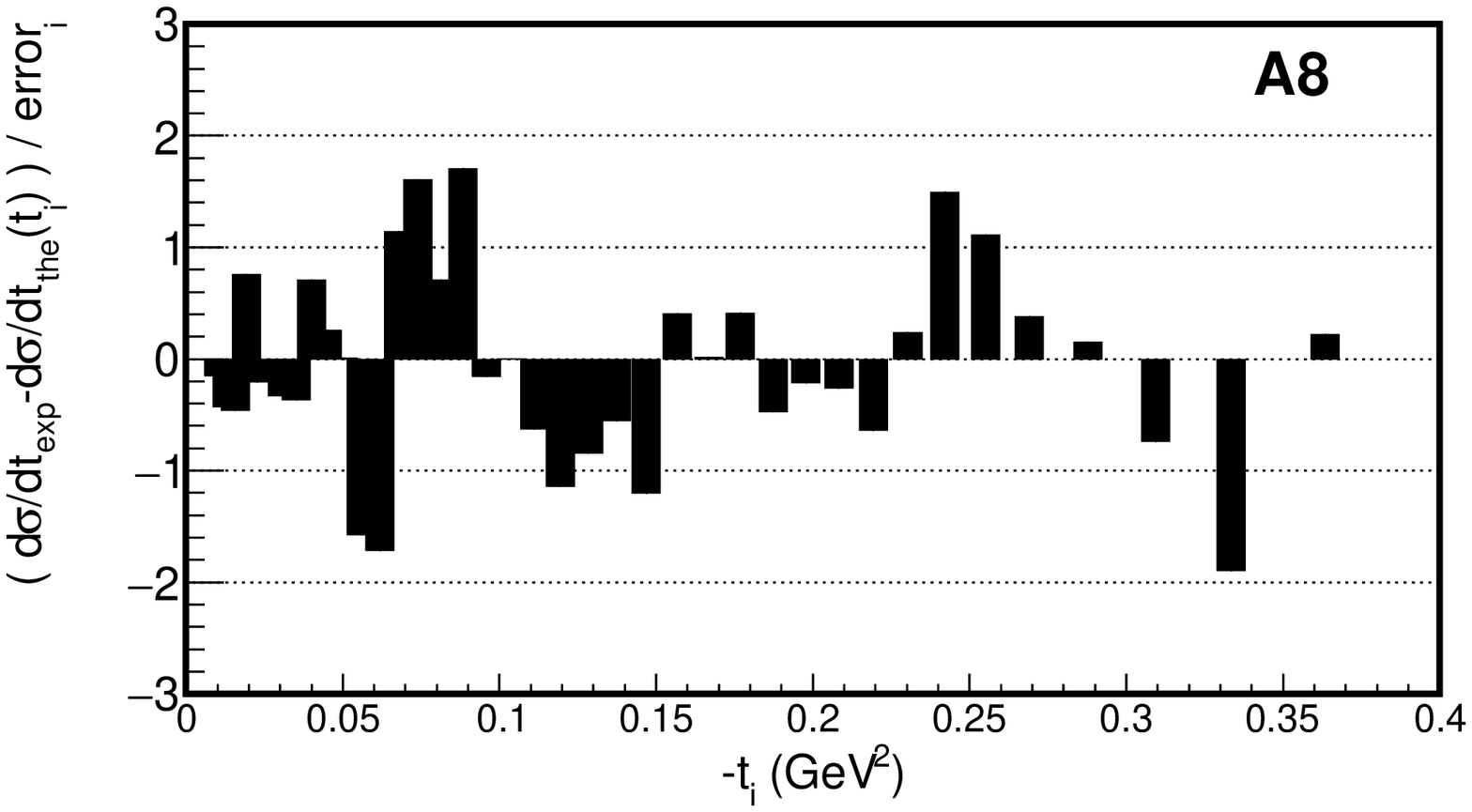}
\includegraphics[width=8cm]{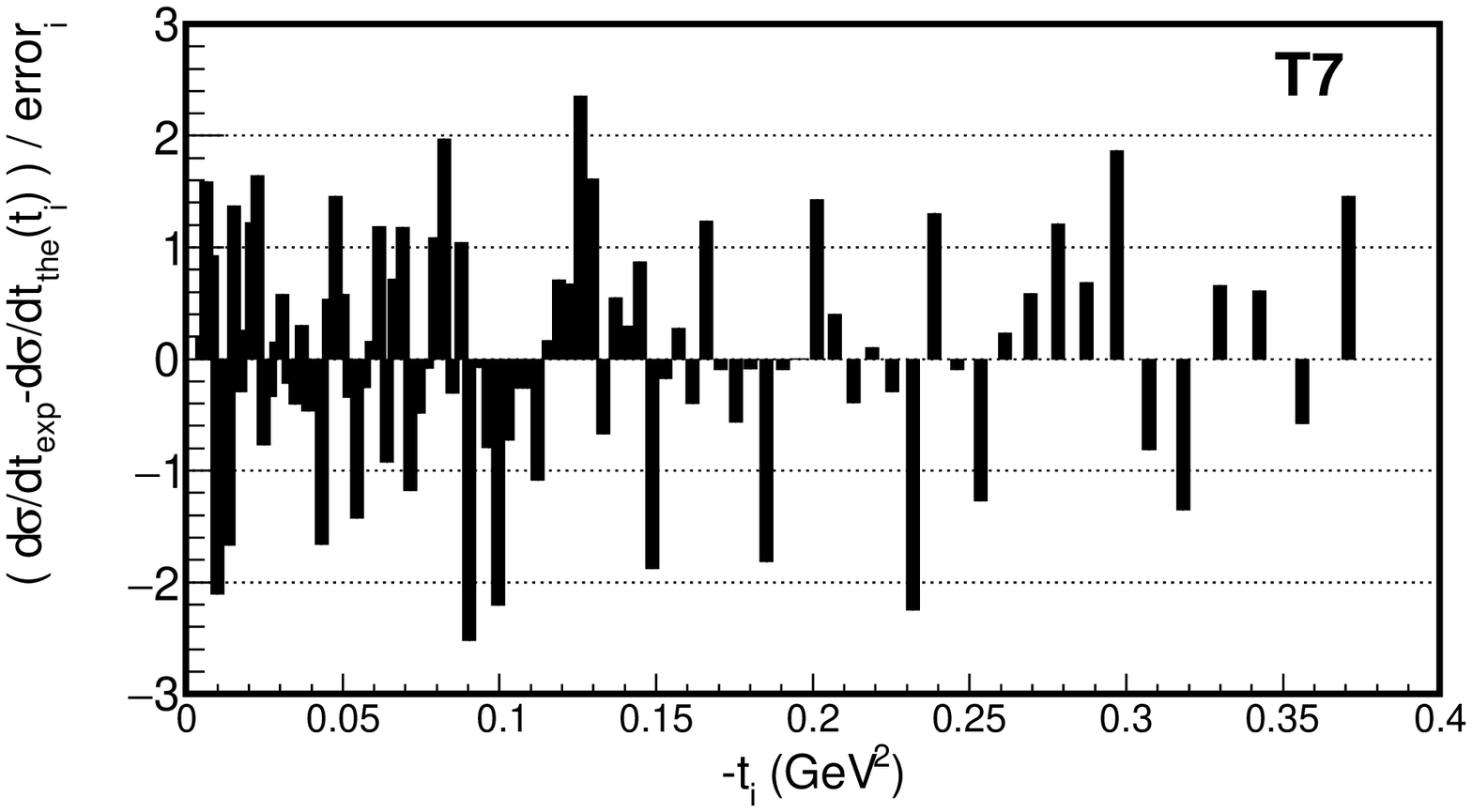}  
  \includegraphics[width=8cm]{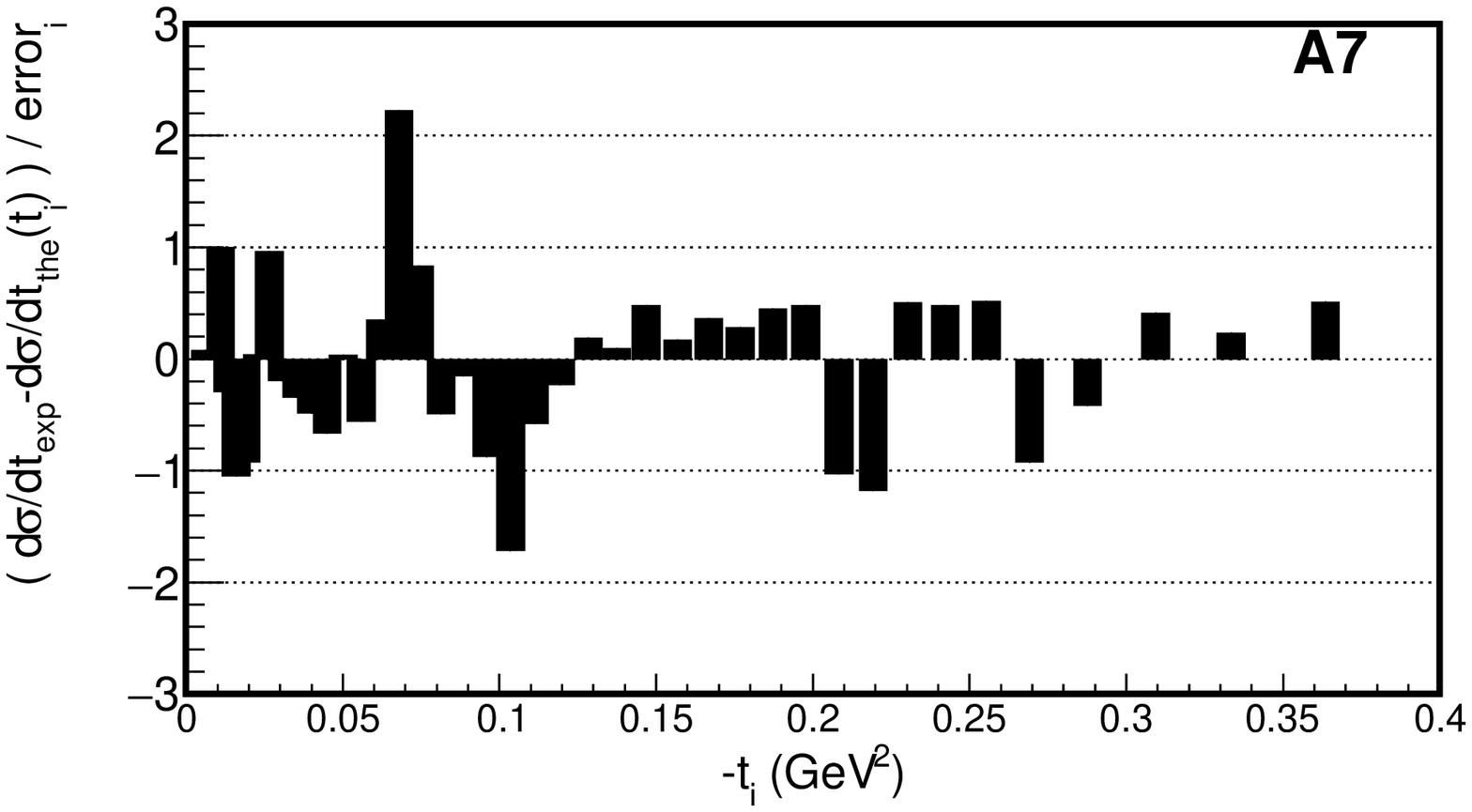}  
\caption{Pull plots of the analysis of the measurements. The y axis in a pull plot is 
defined as $( d\sigma/dt_{\rm exp}^{i}$-$d\sigma/dt_{\rm the}(t_i) )/ {\rm error}_i $,  
where $d\sigma/dt_{\rm exp}^{i} $ is the experimental value at some $t_i$ with ${\rm error}_i $ and 
$d\sigma/dt_{\rm the}(t_i)$ is the theoretical value calculated at $t_i$. According to the maximum 
likelihood criterion (assuming statistical errors following Gaussian distributions) the best 
descriptions of the data are concentrated in T8 and T7 experiments where about 68 $\%$ of the 
experimental points are within 1 sigma deviation from the model.}
\label{pull} \end{figure*}

 The quality of the representations of the data can be read from the pull plots in 
Fig. \ref{pull}. The y axis  represents the standard deviations at each $|t|$ and is 
defined as $( d\sigma/dt_{\rm exp}^{i}$-$d\sigma/dt_{\rm the}(t_i) )/ {\rm er}_i$ ,  
where $d\sigma/dt_{\rm exp}^{i} $ is the experimental value at some $t_i$ with 
error ${\rm er}_i $, and $d\sigma/dt_{\rm the}(t_i)$ is the theoretical value calculated 
at $t_i$. Assuming that the statistical errors follow Gaussian distributions, the 
most probable solution should contain about 68$\%$ of the points within $\pm 1 \sigma$ 
(deviation) and about 95$\%$ of the points within $\pm 2 \sigma$. For T8 
we see that about 65$\%$ of the points are within $\pm 1 \sigma$ following this 
 criterion, and for A8 about 74$\%$ of the points are within  $\pm 1 \sigma$. 
Of course care should be taken in this analysis because  a large number 
of experimental points are needed for a good statistics.
 
 For T7, about 65$\%$ of the experimental points are within $\pm 1 \sigma$ 
and  about 94$\%$   within $\pm 2 \sigma$ of deviation with respect to the 
theoretical curve (fitted curve). Since these are the measurements with larger number 
of points (N=87),  the maximum likelihood criterion for Gaussian statistical errors 
shows that our curve is a good representation of the data.

  The regularity on the values of $\mu_R$ is remarkable. 
 The zero of the real part $t_R$ determined by the parameters $\mu_R$ and $\rho$ is 
associated with the predicted zero of the theorem by A. Martin \cite{Martin}. We 
see that the position  of the zero is stable in all experiments and 
in Table \ref{Table: FINAL} we observe   agreement  
at $t_R\simeq -0.037$ GeV$^{2}$ within the statistical errors. In terms of 
amplitudes we observe that the position of the zero together with the magnitude 
of $B_R$ determines the structure shown in  Fig. \ref{displacement}. The existence 
of this zero is very important for the superposition of the real and imaginary parts
that controls the detailed structure of $d\sigma/dt$ in the low $|t$ region.  

Finally, as a general remark we observe that  total cross sections maintain  that Totem 
values are higher than Atlas  by 4-6 mb, and the numbers remain stable in all 
Conditions studied. Another general remark is that differences in the results 
obtained with the two assumptions for the Coulomb phase are relevant only for T8
that has experimental points for very small $|t|$. 
It  may be fortuitous, but we observe that $\chi^2$  is smaller when the phase is 
put at $\phi_0=0$. 

  The difficulty in the  determination of  $\rho$  may be due 
to the Coulomb-Nuclear interference. The distribution of electric charge  in 
the  proton determined by electromagnetic scattering at low energies 
($\Lambda^2=0.71~\GeV^2$)   may be not realistic for high energy hadronic scattering.  
 In Appendix A we show that with expanded proton size (as it may be the case 
at high energies) the Coulomb phase decreases. 
 We see that the problem of the Coulomb interference 
in elastic pp scattering still  has   open questions.  


 \clearpage

\section{Conclusions   }

In this work we study the properties of the amplitudes in  pp elastic scattering 
analysing experimental data at the LHC center-of-mass energies 7 and 8 TeV, 
  based on a model for the 
complex amplitude, with explicit real and imaginary parts, each containing an 
exponential slope and a linear factor to account for the existence of a zero. 
The zero of the real part, 
close to the origin, corresponds to Martin's Theorem, and  the zero of the imaginary 
part  anticipates the dip in the differential cross section that occurs beyond 
the range of the available data under study.

Our study shows  that the real amplitude plays crucial role in the description of the 
differential cross section in the forward region.
Interference with the Coulomb interaction is properly accounted for, 
and use is made of  information from  external sources, such as dispersion relations 
and predictions for the imaginary zero obtained in studies of full-t behaviour 
of the differential cross section \cite{LHC8TeV,LHC7TeV}. 
  We organize the analysis under four conditions,
according to the specifications of the parameters with values fixed in each case. 
Comparison is made of the results obtained for  the four experimental 
measurements. We obtain the results shown in Table~\ref{Table: FINAL} that we believe to be 
a good representation of the experimental  data of Table \ref{datasets}. 
\begin{table*}[t]
\begin{center}
 \vspace{0.5cm}
 \small
\begin{tabular}{c c  c c c c c c c c c c c}
\hline
\hline
           &     &   & &  &   &         &   &  &  &   &         \\
\multicolumn{12}{c}{Fixed Quantities :   $\rho = 0.14 $ , $\mu_I = - 2.16 \GeV^{-2}$ (8 TeV) \cite{LHC8TeV}, $\mu_I = -2.14 \GeV^{-2}$ (7 TeV) \cite{LHC7TeV} }  \\
\hline 
\hline 
&   N &$\sigma$ & $\rho$        &$ B_{\rm I}  $      &$ B_{\rm R}  $      & $~\mu_{\rm R}  $     & $ \mu_{\rm I} $  & $t_R$  & $B_I^{\rm eff}$  & $B_R^{\rm eff}$ &  $\chi^2/$ndf\\
  &         & (mb)                &$ $ &$(\GeV^{-2})$ &$(\GeV^{-2})$ & $ (\GeV^{-2})$& $ (\GeV^{-2})$ & $(\GeV^{2})$ & $(\GeV^{-2})$& $(\GeV^{-2})$ &         \\
      \hline
       \hline
T8&   60   &102.40$\pm$0.15& 0.14 (fix)& 15.27$\pm$0.39      &21.15$\pm$0.39      & -3.69$\pm$0.15     & -2.16 (fix) &   -0.038$\pm$0.002 &19.59$\pm$0.39 & 73.86$\pm$2.18 & 69.2/56  \\
  \hline
A8&  39   &96.82$\pm$0.11& 0.14 (fix) & 15.26$\pm$0.06      &21.65$\pm$0.24      & -3.69$\pm$0.12     & -2.16 (fix) &  -0.038$\pm$0.001 &20.03$\pm$0.14 & 74.36$\pm$1.73 & 29.97/35  \\
 \hline
T7&   87    &99.80$\pm$0.21 &0.14 (fix) & 15.71$\pm$0.14      &24.26$\pm$0.47      & -4.24$\pm$0.31     & -2.14 (fix) & -0.033$\pm$0.002 &19.99$\pm$0.14 & 84.83$\pm$4.45 & 95.08/83  \\
\hline
T7& 87+17  &99.44$\pm$0.14 &0.14 (fix) & 15.44$\pm$0.07      &22.62$\pm$0.19      & -3.49$\pm$0.13     & -2.14 (fix)& -0.040$\pm$0.002 & 19.72$\pm$0.07 & 72.48$\pm$1.87 & 203.5/100  \\
\hline
A7& 40   &95.75$\pm$0.16&0.14 (fix) & 15.23$\pm$0.11      &21.86$\pm$0.44      & -3.99$\pm$0.22     & -2.14 (fix) & -0.035$\pm$0.002 &19.51$\pm$0.11& 78.86$\pm$3.17 & 27.33/36  \\
\hline
\hline
\end{tabular}
 \caption{Proposed values of parameters for the  four datasets. The phase of the 
Coulomb interference is calculated  with electromagnetic proton form factor as 
described in Appendix A. The T7 data are also shown with inclusion of points at 
higher $|t|$  that are important for confirmation of the value of $\mu_I$.   }
\label{Table: FINAL}
\end{center}
\end{table*}
 
Assuming $\sigma\sim \log^2{s}$ at high energies, dispersion relations give  
$\rho \sim \pi/\log s$, and we know that  the position of the first real 
zero \cite{LHC8TeV} 
behaves like $|t_R| \simeq A+1/(c_0+c_1\log{s}+c_2\log^2{s})$. From Eq.(\ref{tR_position}) 
it then  follows that $\mu_R$ increases like   $ - \lambda \log{s}$ at high energies.
With $s$ in $\GeV^2$, from Table \ref{Table: FINAL} we obtain     
 $\lambda \simeq 0.2 ~\GeV^{-2}$. 
Future precise measurements in LHC at 13 TeV  may be investigated with this purpose. 
Our present work predicts that the zero at $\sqrt{s}= 13 $ TeV is at  $t_R=-0.037 \GeV^2$.

The   quantity  $\mu_R$   is related  with the scaling variable $\tau = t \log^2{s}$ 
introduced by J. Dias de Deus   \cite{Deus} connecting $s$ and $t$ dependences 
in the amplitudes  at high energies and small $|t|$.
 A. Martin \cite{Martin_Real}  uses the same idea of a scaling variable,  writing 
an equation for the real part $\rho(s,t)$ using crossing symmetric scattering 
amplitudes   of a complex $s$ variable,  valid in a forward range. 
 The proposed ratio  is 
\begin{equation}
\rho(s,t)\simeq \frac{\pi}{\log s}\Big(1+\frac{\tau(df(\tau)/d\tau)}{f(\tau)}\Big) ~ ,  
\label{real_martin}
\end{equation}
where $f(\tau)$ is a damping function, with the implicit existence of a real zero.
The form  of $f(\tau)$ determines the properties of the real zero  \cite{Dremin}. 
that is found in the analysis of the data. 
This may be a clue for the introduction of explicit crossing symmetry and   
analyticity in our phenomenological treatment of the data.

Other models \cite{Models} also deal with the position of the real zero, 
discussing different analytical forms for the amplitudes, and it would 
be interesting to have their predictions for the amplitudes in the 
forward range.

In Appendix B  we investigate the behaviour the  amplitudes in  $b$-space for  
a geometric insight of  the scattering process. We find that although our  amplitudes 
  never violate  Eq.(\ref{eldominant}) up to $\sqrt{s} = $ 8 TeV at LHC, the differential 
cross sections in $b$-space are all monotonically decreasing functions in $b$. 
This means that our amplitudes do not indicate any signal for   peripheral dominant 
elastic scattering process. However, in an analysis of T8  
a dominance of the square  of the real amplitude in $b$-space    Ref. \cite{Jiri}       
interpreted as a peripheral behaviour. To settle this interesting question,  
  data at larger $|t|$ are required. 
 
In non-perturbative QCD, in several instances, the proton  appears as 
a structure with expanding size as the energy increases \cite{sizes}, 
with varied mechanisms, as distribution of valence quarks 
in a cloud around a core, modifications in QCD vacuum in the region of the 
colliding particles, and so on. 
Together with the evolution of the proton hadronic size, its electromagnetic 
properties, as they appear in high energy collisions, may change also. 
A linear increase in $\log{s}$ is a usual assumption for the effective proton radius,  
and   the form factor parameter $\Lambda^2$ would  then be reduced by about 1/2, 
corresponding to increase of   about 40 \% in proton radius. In Appendix A we calculate the 
interference phase with this example.

We expect that future data in pp elastic scattering at 13 TeV 
and higher energies will have high quality covering a wide $t$ range 
to allow  determination of the properties  of the real and imaginary 
amplitudes in pp elastic scattering, including studies of 
the   amplitudes up to the perturbative tail of large $|t|$.  
Hopefully the experimental groups will receive  the necessary  support and 
encouragement for this effort.

\clearpage

 \begin{acknowledgments}
TK, EF and AKK thank INCT-FNA Proc.No. 464898/2014-5 for the support. The authors wish to thank the Brazilian agencies CNPq, CAPES, PRONEX and FAPERJ for financial support. AKK thanks  Jiri Prochazka for stimulating discussions during EDSBlois 2017.
\end{acknowledgments}

\appendix 
\section  {Coulomb Phase}\label{AppendixA}

{\bf Abstract} 
We study the phase of  t
scattering according to the formulation based on superposition of nuclear and 
Coulomb interactions in the eikonal formalism. The expressions for the phase accounting 
for the  structure of the real and imaginary parts of the amplitude are obtained. 

\bigskip

The   Coulomb interference in pp elastic scattering is very important for the 
analysis of   LHC experiments, where the properties of the amplitudes are investigated. 
The phase   $\phi$  of the  quantum interference   between Coulomb and nuclear interactions was
studied   by  several authors  \cite{WY,Cahn,LHC7TeV,KL,Selyugin,Petrov2}.  
    A first approach for the 
so called Coulomb phase,  is the point-like (pure Coulomb) phase, derived by West Yennie \cite{WY} 
calculated in the  context of quantum field theory,   considering  the nuclear 
imaginary and real amplitudes with the same t-dependence and a proton with zero 
radius.  The  results for pointlike proton \cite{WY} have been  generalized 
\cite{LHC7TeV} allowing different slopes  for the real and imaginary nuclear amplitudes. 
A second approach is the form-factor phase, which considers an eikonal 
treatment for the superposition of amplitudes and a geometrical size with a form factor for the 
proton structure \cite{Cahn,KL,Selyugin}.

The expression for the phase depends on   
the analytical structure of the real and imaginary parts of the nuclear amplitude.
Here we derive in the same eikonal formalism  an expression for the Coulomb interference 
phase appropriate for forward scattering amplitude  with $t$ dependence of the form   
\begin{eqnarray}
&&  T(t)=   \big(\rho_R-\mu_R t -\xi_R t^2\big)e^{B_R t/2}  \nonumber \\
&& +i \big(\rho_I-\mu_I t -\xi_I t^2    \big)e^{B_I t/2}  ~, 
\label{Nuclear-ampl} 
  \end{eqnarray}
that is assumed to be realistic   for the description of $d\sigma/dt$ data for high 
energy,  satisfying 
expected properties concerning zeros, signs and magnitudes of the amplitudes 
in the forward range.  
   To have systematic symmetry in the notation, we write $\rho_R=\rho$ for 
the usual,$\rho$ parameter, and $\rho_I=1$ . 
 
 The  usual  dipole  electromagnetic form factor is 
 \begin{equation}
f(q^2)=\Big(\frac{\Lambda^2}{q^2+\Lambda^2}\Big)^2~,
\label{form_factor_0} 
 \end{equation}
with $\Lambda^2=0.71 \GeV^2$. To simplify calculations,  
~we may alternatively use the   
   Gaussian representation for  the form factor
 \begin{equation}
f(q^2)=e^{-2q^2/\Lambda^2}~, 
\label{form_factor_1} 
 \end{equation}     
trusting that  differences in results for the phases  are unimportant, as confirmed by 
Cahn \cite{Cahn}.

The Coulomb amplitude for pp scattering, written
\begin{equation}
F^C(s,q^2)=-\frac{\alpha s}{(q^2)} f^2(q^2) 
\label{Coulomb_ampl}
\end{equation}
  enters the Coulomb-Nuclear-Interference (CNI)  
in the form
\begin{equation}
F^{N+C}(s,q^2)=-\frac{\alpha s}{(q^2)} f^2(q^2) ~ e^{i\alpha \phi}  + F^N(s,q^2) 
\label{Coulomb_ampl_2}
\end{equation}
with normalization defined by 
\begin{eqnarray}
&&  \sigma=\frac{4 \pi}{s} Im F^N(q=0)    ~ , \nonumber \\
&& - \frac{d \sigma}{dq^2}= \frac{\pi}{s^2} |F|^2 ~ . 
\label{normalization}
\end{eqnarray}

 We start from the exact expression for the phase
\begin{eqnarray}
&&\phi (s,-q^2)=\nonumber \\
&&\mp \int_0^{\infty}d^2{\bf{q'}}\ln\Big(\frac{q'^2}{q^2}\Big)\frac{d}{dq'^2}
\Big[f^2(q'^2)\frac{F_N(s,[{\bf{q}}-{\bf{q'}}]^2)}{F_N(s,q^2)}\Big]~, \nonumber \\
\label{Cahn_phase_2}
\end{eqnarray}
the signs  $\mp$ corresponding to the  $ {\rm pp/p\bar p } $ systems.

Cahn has considered   the approximation 
\begin{eqnarray}
&& (1/(2\pi)) \int_0^{2\pi}d\theta  F_N(s,({\bf q'-q})^2)/F_N(s,q^2) \nonumber  \\
&& \approx F_N(s,(q'^2)/F_N(s,0) 
\label{approx}
\end{eqnarray}
justifying 
that the momentum transfer scale $q^2$ is much smaller than the inverse 
size of the proton. This is a very strong assumption, that should be tested 
in specific cases.  
Following the suggestion, Eq. (\ref{Cahn_phase_2}) becomes 
\begin{eqnarray}
\phi (s,t)=\mp \int_0^{\infty}dt'\ln\Big(\frac{t'}{t}\Big)\frac{d}{dt'}
\Big[f^2(t')\frac{F_N(s, t')}{F_N(s,0)}\Big]~, \nonumber \\
\label{Cahn_phase_app}
\end{eqnarray}
where for simplicity we use  the $t'$ variable instead of $q'^2$.  
As a generalization with respect to Cahn's calculation, we take for the nuclear amplitude 
the same expression in Eq. (\ref{Nuclear-ampl}), 
  namely $ F_{N}(s,t)=$ 
$$ =  \sigma(\rho-\mu_R~t-\xi_R~t^2)e^{B_{R}~t/2}+i\sigma(1-\mu_I~t
      -\xi_I t^2)e^{B_{I}~t/2}~.$$ 
Then we  need to evaluate integrals 
\begin{eqnarray} 
&& H_N(t,b_\beta)=  \int_0^{-\infty}dt'\ln\Big(\frac{t'}{t}\Big)~
\frac{d}{dt'}\big[t'^N ~ e^{4t'/\Lambda^2} e^{B_\beta t'/2}  \big] ~ \nonumber \\
&& = \int_0^{-\infty}dt'\ln\Big(\frac{t'}{t}\Big)~\frac{d}{dt'}\big[t'^N ~ e^{b_\beta t'}\big]~ ,  
\label{integral_N} 
\end{eqnarray} 
where we have used the definition
$$b_\beta= \frac{4}{\Lambda^2} +\frac{B_\beta}{2} $$ 
with $\beta= R, I $.  

The results of the integrations (N=0,1,2) are 
\begin{eqnarray}
&&  H_0=\gamma ~+ ~\log(- b_\beta t) ~ , \nonumber \\
&&  H_1= \frac{1}{b_\beta} ~ ,  \nonumber \\
&&  H_2= - \frac{1}{b_\beta^2} ~,  
\end{eqnarray} 
where $\gamma=0.5772$  is the Euler Gamma constant.
The  phase is then  written 
\begin{eqnarray}
&& \phi (s,t)=\mp \frac{1}{\rho + i} \big\{ \big[- \frac{\mu_R}{b_R}+\frac{\xi_R}{b_R^2}  
 + \rho \big(\gamma + \log(- b_R t)\big) \big]  \nonumber  \\
&& + i ~  \big[- \frac{\mu_I}{b_I}+\frac{\xi_I}{b_I^2}   + \gamma + \log(- b_I t) \big]   \big\} ~ ,  
\label{our_phase_1}
\end{eqnarray}
with real and imaginary parts respectively 
 \begin{eqnarray}
&& \phi_{R} (t)=\mp\bigg\{\frac{1}{1+\rho^2}  
  \bigg[  \big[- \frac{\mu_I}{b_I}+\frac{\xi_I}{b_I^2}+\log(b_I) \big] \nonumber \\
&& + \rho \big[ - \frac{\mu_R}{b_R}+\frac{\xi_R}{b_R^2} 
       +\rho ~ \log(b_R)\big] \bigg]+ \gamma + \log[-t]  \bigg\}
\label{Cahn_phase_real_2}
\end{eqnarray} 
and  
\begin{eqnarray}
&& \phi_{I} (t)=   \mp \frac{1}{1+\rho^2} 
   \bigg\{ \rho  \big[- \frac{\mu_I}{b_I}+\frac{\xi_I}{b_I^2}+\log( b_I) \big] \nonumber \\
&& - \big[- \frac{\mu_R}{b_R}+\frac{\xi_R}{b_R^2}   + \rho \log( b_R ) \big] \bigg\} ~. 
\label{Cahn_phase_imag_2}
\end{eqnarray} 
 Eqs. (\ref{Cahn_phase_real_2}) and (\ref{Cahn_phase_imag_2})   are our final results for the 
phase calculated with form factors,  in a generalization of the work by Cahn \cite{Cahn},  
assuming more complete structures for the real and imaginary parts of the scattering amplitude. 
   It may be of practical usefulness  to define  
\begin{equation}
  C_R =   - \frac{\mu_R}{b_R}+\frac{\xi_R}{b_R^2}   + \rho ~ \log( b_R ) 
\label{C_real}
\end{equation}
and 
\begin{equation}
  C_I =   -\frac{\mu_I}{b_I}+\frac{\xi_I}{b_I^2}  +\log( b_I )  ~.     
  \label{C_imag}
\end{equation}
  and then write     
\begin{equation}
 \phi_{R} (t)= \mp  \bigg[     \frac{1}{1+\rho^2}     
 \big[ C_I+ \rho ~ C_R  \big]   +  \gamma + \log(-t) \bigg]  
\label{final_R}
\end{equation}
and 
\begin{equation}
 \phi_{I} (t)= \mp \frac{1}{1+\rho^2}\big[\rho ~ C_I-C_R \big]    ~. 
\label{final_I}
\end{equation}

It must be observed that in these expressions  $b_R$ , $b_I$   and $-t$  have 
compatible units, as $\GeV^{-2}$  and $\GeV^2$. 
The result is simple: in the real part the $t$ dependence is purely linear in $\log(-t)$, 
the imaginary part is very small  constant, and there is no explicit energy dependence. 

In the simplified case  
$\mu_R=\mu_I=0, ~  \xi_R=\xi_I=0,  ~ B_R=B_I=B, $
$$ b_R=b_I=b=\frac{4}{\Lambda^2}+\frac{B}{2} $$ 
we obtain  Cahn's original form.
 
\begin{figure}
                \includegraphics[width=8.5cm]{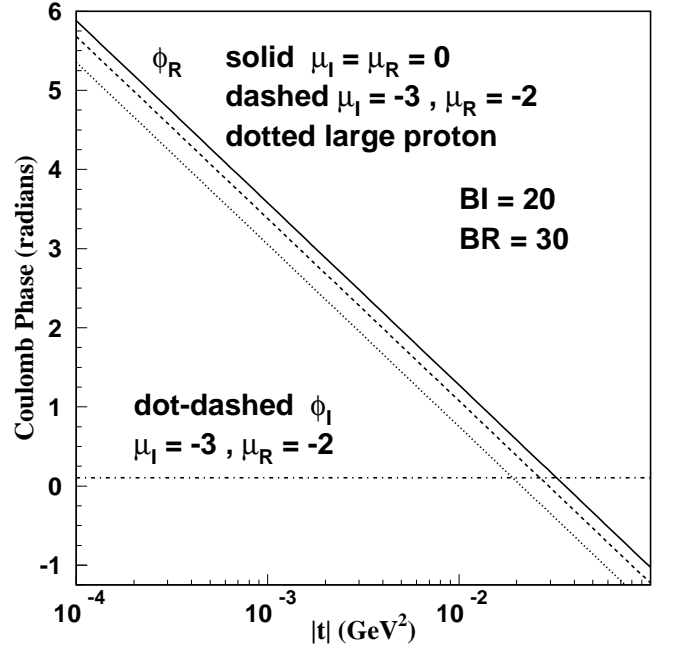}
   { \caption  {Phase of the Coulomb-Nuclear interference with examples for 
the values of the parameters.
We plot also the phase calculated with $\Lambda^2=0.71/2 \GeV^2$  } }
             \label{phase}
     \end{figure}

  The real and imaginary parts of the phase are plotted in 
Fig. \ref{phase} with an example of values for the parameters. 

In the figure we plot also (dot-dashed) a line representing the real part of the phase
calculated for a larger proton, with $\Lambda^2=0.305 \GeV^2 $. 

\clearpage 

\section{$ b$-space properties \label{bspace}}
 
\begin{figure}[b]
                \includegraphics[width=7 cm]{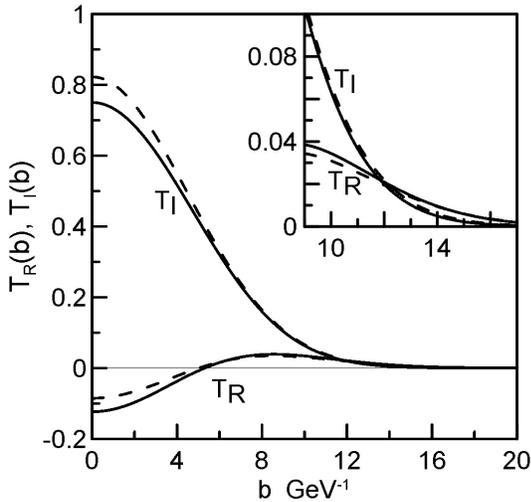}
    \caption  {  $b$-space amplitudes at 8 TeV. Solid and dashed correspond to 
A8 and T8 measurements respectively. As seen the
inset in the right-up corner,  real parts cross the imaginary parts around
$b\sim12$ ${\rm GeV}^{-1}$ and become dominant for the peripheral region. The 
structures for 7 TeV are  similar. 
 }
             \label{FigAmp8}
     \end{figure}

Let $\tilde{T}(b;s)$ be the dimensionless Fourier transform of the amplitude $T(s,t)$ in 
Eqs. (\ref{imag_TI}, \ref{real_TR}) 
 with respect to the momentum transfer. Writing  
\begin{equation}
\tilde{T}(b;s)=\tilde{T}_{R}+i\tilde{T}_{I},
\end{equation}
we have 
\begin{eqnarray}
\tilde{T}_{R}\left(  b;s\right)  =\frac{\sigma}{2\pi B_{R}}\Bigg\{
\rho+\frac{\mu_{R}}{B_{R}}\left(  2-\frac{b^{2}}{B_{R}}\right) \Bigg\}  e^{-b^{2}/2B_{R}} ~, 
\end{eqnarray}
and 
\begin{eqnarray}
\tilde{T}_{I}\left(  b;s\right)     =\frac{\sigma}{2\pi B_{I}}\left\{
1+\frac{\mu_{R}}{B_{R}}\left(  2-\frac{b^{2}}{B_{I}}\right)  \right\}
e^{-b^{2}/2B_{I}}.
\end{eqnarray}
  In Fig.(\ref{FigAmp8})   we show
    $\tilde{T}$ at $\sqrt{s}$= 8 TeV as function of $b$ for T8 and A8.  

In Fig. (\ref{FigAmp8}), we show  the general tendency of the scattering
amplitudes in the two measurements, that differ by only a few \%. 
Note that the magnitude of real parts become  comparable,
and even greater than the imaginary parts for large $b$ values. Such a
behaviour is not the case for lower energies. The similar behaviour is similar 
for  the 7  TeV datasets. 

To investigate in more detail the significance of the behaviour of the real
part dominance in the peripheral region, let us introduce the {\it eikonal
representation} of the $b-$ space amplitude as (\cite{LHC8TeV})
\begin{equation}
\tilde{T}\left(  b,s\right)  \equiv i\sqrt{\pi}\left(  1-e^{i\chi\left(
s,b\right)  }\right)  ,\label{Tb}%
\end{equation}
and the so-called $b-$ space differential cross sections (profile functions) are 
\begin{equation}
\frac{d\tilde{\sigma}_{\rm el}\left(  s,\vec{b}\right)  }{d^{2}\vec{b}}%
=1-2\cos\chi_{R}e^{-\chi_{I}}+e^{-2\chi_{I}},\label{el}%
\end{equation}%
\begin{equation}
\frac{d\tilde{\sigma}_{\rm tot}\left(  s,\vec{b}\right)  }{d^{2}\vec{b}}=2\left(
1-\cos\chi_{R}e^{-\chi_{I}}\right)  ,\label{tot}%
\end{equation}%
\begin{equation}
\frac{d\tilde{\sigma}_{\rm inel}\left(  s,\vec{b}\right)  }{d^{2}\vec{b}%
}=1-e^{-2\chi_{I}},\label{inel}%
\end{equation}
where $\chi_{R}=\operatorname{Re}\left(  \chi\right)  $ and $\chi
_{I}=\operatorname{Im}\left(  \chi\right)$. 

 The above $b$-space representations of the differential cross sections 
  offer a geometric view of the pp interactions,
although such interpretation should be taken with care because they are 
not physical observables. 
From the unitarity condition of the scattering amplitude, we must 
have $\chi_I\ge 0$, and $d\sigma_{\rm inel}/d^2 b \le1$. 
We have $\left\vert \tilde{T}\right\vert \ll1$ which implies $\left\vert
\chi\right\vert \ll1$. Thus, in this region,  
 taking up to the leading orders in
$\chi_{R}$ and $\chi_{I}$ respectively, we  obtain 
\begin{equation}
\frac{d\tilde{\sigma}_{\rm el}\left(  s,\vec{b}\right)  }{d^{2}\vec{b}}\sim
\chi_{_{R}}^{2}=\frac{1}{\pi}\tilde{T}_{R}^{2} ~ ,\ \
\end{equation}%
\begin{equation}
\frac{d\tilde{\sigma}_{\rm tot}\left(  s,\vec{b}\right)  }{d^{2}\vec{b}}%
\sim2\left(  \chi_{I}+\frac{1}{2}\chi_{_{R}}^{2}\right)  =\frac{2}{\sqrt{\pi}%
}\tilde {T}_{I}+\frac{1}{\pi}\tilde{T}_{R}^{2}%
\end{equation}%
and 
\begin{equation}
\frac{d\tilde{\sigma}_{\rm inel}\left(  s,\vec{b}\right)  }{d^{2}\vec{b}}%
\simeq2\chi_{I}\simeq\frac{2}{\sqrt{\pi}}\tilde {T}_{I} ~ .
\end{equation}
From the above, it is clear that in the domain where $\left\vert \tilde
{T}\right\vert \ll1$ (that is, where the Born approximation is valid), the inelastic
part comes totally from $\tilde{T}_{I},$ whereas the elastic contribution may
exceed the inelastic one if
\begin{equation}
\tilde{T}_{R}^{2}>2\sqrt{\pi}\tilde{T}_{I}.\label{eldominant}%
\end{equation}
Thus the appearance of the peripheral domain where
$\tilde{T}_{R}>\tilde{T}_{I}$ may
indicate that the contribution of elastic scattering is going to be
significant in the peripheral region, and  even can be dominant if   relation
(\ref{eldominant}) is satisfied. If this  situation happens, the scattering
is basically elastic and a possible candidate that we can imagine in 
such situation is the elastic scattering
channel due to   virtual pion exchange.    

 In the model of M. M. Islam \cite{Islam} the proton structure is described by three 
layers: the first within a radius of $\sim 0.2$ fm contains three valence quarks 
(exchange of small-x gluons), the second layer is a shell of baryonic charge $\sim 0.44$ fm (responsible for $\omega$ exchanges) and the outer layer is the $q\bar{q}$ condensate which dominates   diffractive scattering and is the same for both proton and anti-proton. However the polarization induced by the inner shells change the distribution of charges around the proton from pp to p\=p. This re-distribution affects the amplitudes and creates a contribution which is energy dependent. In other words this can be interpreted as a change in the proton electromagnetic form factor. 
 On the other hand, in
the present analysis, the relation (\ref{eldominant}) is far from be  satisfied 
for all datasets, for $7$ and $8$ TeV. That is, there appear no elastic scattering
dominant outer shell at these and lower energies. 


Below, we show the plots of various $b$-space differential cross sections
using the amplitudes obtained in the present analysis.
Note that the corresponding $b$-space amplitude satisfies the unitarity condition
mentioned before, satisfying 
$$0\le d\sigma_{\rm el}/db, ~ d\sigma_{\rm inel}/db \le 1 ~.  $$

    \begin{figure}[b]
                \includegraphics[width=7 cm]{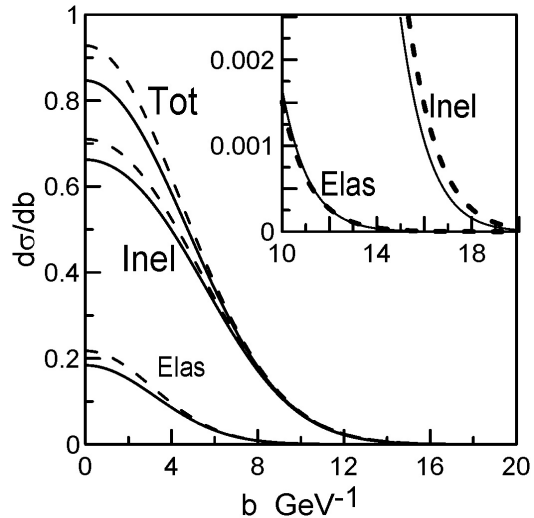}
    \caption  {    $b$-space differential cross sections for $\sqrt{s}$ = 8 TeV. Solid  and dashed curves for A8 and T8 respectively. In all   $b$ domain, the inelastic contribution is far dominant over the elastic one. }
             \label{dsdb-8TeV}
     \end{figure}

As seen from these figures, for all results,  although the real part amplitude
becomes dominant over the imaginary part in the peripheral  region,  the
inelastic contribution is totally dominant over the elastic contribution at
these energies (7 and 8 TeV). On the other hand, if such a tendency of
increase of the real part continues for much higher energies and eventually
  becomes dominant compared to the imaginary part for the peripheral
region, then the contribution from the elastic contribution becomes larger and
the ratio $\sigma_{\rm el}/\sigma_{\rm tot}$ may increase, approaching the black-disk
limit. If  this  scenario happens and this is due to the elastic scattering by 
pion cloud, then we would expect that the Froissart bound might be saturated. 
However, as far as the data indicate the above speculative scenario where 
the real part becomes dominant, it is very unlikely for few hundred TeV energies, 
even considering the sensibilities and uncertainties in the determination of real and imaginary amplitudes.
The question of asymptotic properties for asymptotic energies, we require more
careful analysis of the real and imaginary amplitudes  only for the forward 
region but  also  for the larger $|t|$ values.


\end{document}